\documentclass{llncs}  

\usepackage{fullpage}

\usepackage{balance}  
\usepackage{graphicx}
\usepackage{url}
\usepackage{cite}

\usepackage{color}

\newcommand{\remove}[1]{}

\usepackage{float}

\begin{document}


\title{Ride Sharing and Dynamic Networks Analysis}

\author{Tal Altshuler$^{1}$ \and Rachel Katoshevski$^{1}$ \and Yoram Shiftan$^1$}
\institute{$^1$Department of Civil and Environmental Engineering, Israel Institute of Technology}

\maketitle

\abstract{
The potential of an efficient ride-sharing scheme to significantly reduce traffic congestion, lower emission level and drivers' stress, as well as facilitating the introduction of \emph{smart cities} has been widely demonstrated in recent years \cite{santi2014quantifying}.
Furthermore, ride sharing can be implemented within a sound economic regime through the involvement of commercial services that creates a win-win for all parties (e.g., \emph{Uber}, \emph{Lyft} or \emph{Sidecar} \cite{uber,lyft,side}).
This positive thrust however is faced with several delaying factors, one of which is the volatility and unpredictability of the potential benefit (or utilization) of ride-sharing at different times, and in different places.
Better understanding of ride-sharing dynamics can help policy makers and urban planners in increase the city's ``ride sharing friendliness'' either by designing new ride-sharing oriented systems, as well as by providing ride-sharing service operators better tools to optimize their services.

In this work the following research questions are posed: (a) Is ride-sharing utilization stable over time or does it undergo significant changes? (b) If ride-sharing utilization is dynamic, can it be correlated with some traceable features of the traffic? and (c) If ride-sharing utilization is dynamic, can it be predicted ahead of time? We analyze a dataset of over 14 Million taxi trips taken in New York City. We propose a dynamic travel network approach for modeling and forecasting the potential ride-sharing utilization over time, showing it to be highly volatile. In order to model the utilization's dynamics we propose a network-centric approach, projecting the aggregated traffic taken from continuous time periods into a feature space comprised of topological features of the network implied by this traffic. This feature space is then used to model the dynamics of ride-sharing utilization over time.

The results of our analysis demonstrate the significant volatility of ride-sharing utilization over time, indicating that any policy, design or plan that would disregard this aspect and chose a static paradigm would undoubtably be either highly inefficient or provide insufficient resources.
We show that using our suggested approach it is possible to model the potential utilization of ride sharing based on the topological properties of the rides network. We also show that using this method the potential utilization can be forecasting a few hours ahead of time.
One anecdotal derivation of the latter is that perfectly guessing the destination of a New York taxi rider becomes nearly three times easier than rolling a ``Snake Eyes'' at a casino.
}

\section{Introduction}

The increasing availability of portable technologies gives new fuel to studies on metropolitan transportation optimization, pushing urban design one step closer towards the long sought concept of ``smart cities'' \cite{caragliu2011smart,chourabi2012understanding}.
Mobile devices and ubiquitous connectivity make it easier than ever to collect data on the way people live in cities and big-data analytic methods facilitate the extraction of actionable insights from it.
City administrators and policy makers can in turn act upon such results to enhance city management, channeling current advancements in data analysis for the immediate improvement of urban quality of life.

Many of the fundamental problems in big cities nowadays relate to cars.
The high number of vehicles congests the streets, vehicles standing in traffic jams increase air pollution while also increasing traveling times, significantly increasing passengers' stress levels.
Availability of large-scale datasets accompanied with recent advancements in the analysis of big-data and the development of novel models of human mobility, give rise to new possibilities to study urban mobility.

Such new models include for example the work of~\cite{gonzalez2008understanding} in which large-scaler mobile phone data was analyzed in order to characterize individual mobility, showing that human travel patterns are far from random, and are efficiently describable by a single spatial probability distribution.
Similarly, \cite{calabrese2013understanding} show that mobile phone data can be used as a proxy to examine urban mobility and ~\cite{noulas2012tale} analyzes social network data of different cities to find that mobility highly correlates with the distribution of urban points of interest.
Mobile technologies are also the enablers of many successful consumer applications, such as Waze~\cite{waze}, that provide traffic-aware city navigation by using data provided by the community.
Alternative ways of moving in the city, such as autonomous mobility-on-demand and short-term car rental have been identified among the possible solutions to the ever-growing transport challenge~\cite{bbcFutureTraffic}.

Ride sharing has the potential of improving traffic conditions by reducing the number of vehicles on the roads, reducing the emission of $CO_2$ and the fuel consumption per person and giving the riders the opportunity to socialize with people (that otherwise would have been fierce ``road competitors'').
A recent study~\cite{cici2014assessing} shows that traffic in the city of Madrid can be reduced by 59\% if people are willing to share their home-work commute ride with neighbors. Even if they are not willing to ride with strangers, but only with friends of friends (for safety issues), the potential reduction is still up to 31\%. Another recent study~\cite{altshulerNetsci} had shown that on-demand route-free public transportation based on mobile phones outperforms standard fix-route assignment methods when comparing traveling times.
These results encourage the deployment and policies supporting ride sharing in urban settings.

However, despite such evidences and others, ride-sharing adoption rate in cities worldwide is slower than what can be expected given the clear benefits of ride-sharing \cite{graziotin2013analysis,alexander2015assessing}. One important reason, as suggested by \cite{hobrink2014explaining,coenen2012toward,shaheen2009north} and others is the uneven, and often unstable, potential benefits associated with ride-sharing. When the value that can be extracted from using a service such as Lyft \cite{lyft} is high at one part of the city, but significantly lower at another neighborhood, or worse -- suddenly decreases for a period of two days -- potential users of the service are much likely to opt for a private car usage \cite{efthymiou2013factors}.

In this work we propose a data-driven framework to dynamically predict the impact, or \emph{potential utilization}, of ride sharing in a city, at different times, and in different regions. Specifically, the technique we propose provides both policy makers as well as ride-charing operators tools for assessing the future benefit of ride-sharing, encapsulated through the percent of rides saveable through merging of nearby departures and destinations. Simply put, a shared taxi service can use this proposed technique in order to know ahead of time what the ride-sharing demand is going to be (at various places in the city), whereas municipal services can dynamically change tolls and service fees in order to incentivize the use of ride-sharing in ``low hours'' that are predicted in advance.

Our method is based on analyzing the network features of the dynamic O-D matrix as represented by data collected by various sources, such as mobile phone call records, or sensors mounted on the taxis themselves. In our research we show a clear correlation between such properties and the portion of ``merge-able rides''.
We have analyzed the efficacy of our proposed network-oriented method using a dataset of over 14 Million taxi trips taken
in New York City during January 2013 \cite{nyc}.

This work is structured as follows: Section~\ref{sec:related} presents an overview of the relevant related research in the field. \remove{In Section \ref{sec:related2} we illustrate our claim that new technologies can be harnessed for the enablement of innovative solutions that rely on policies and economic incentives rather than investments in infrastructures by examining the success of Carsharing services (such as Zipcar \cite{lit_zip1} or City CarShare \cite{LIT33}) that was made possible using the emerging technologies of the 90s.}
In Section \ref{sec:data} we discuss the data and analytic methodologies that were used for this work: starting with the calculation of the average ride-sharing potential as a function of the maximum delay a taxi-user would be willing to sustain, we demonstrate that more than 70\% of the rides can be shared when users are willing to undertake up to 5 minutes delay.
We then demonstrate that urban ridesharing potential is not only highly dynamic, but that it can also be predicted using the analysis of the rides that took place in the city a few hour beforehand. We present a method for comprising a dynamically changing network using the taxi-rides, and analyzing the topological properties of this network (Section \ref{sec:network}). We analyze the dynamics of these properties over time, and demonstrate our ability to accurately predict changes in the utilization of ride-sharing several hours in advance.
Concluding remarks and suggestions for future works are contained in Section \ref{sec:summary_future}.

\remove{
\section{Shared Private Transportation: Feasibility and Evaluation}
\label{sec:related2}

\input{lit2.tex}
}

\section{Related Work}
\label{sec:related}

In this section we bring forth a comprehensive review of the literature relevant to our work. As our suggested methodology is based on the usage of network analysis for a predictive modeling of ride-sharing efficiency, the body of work discussed in this section encompasses a relatively wide variety of works. We start by providing a background review of traffic design approaches used as a component of the urban design process.
We then proceed to review data-driven mobility modeling works -- a specific approach to urban design, focusing on the usage of large-scale datasets for the modeling of human mobility. The importance of such modeling stems from the fact that a better understanding of ``how people travel'' that is based solely on data (in contrast to various psychological or economic approaches) is in many cases proven to be more accurate, flexible and robust. Such research thrusts however have only recently been made possible, with the appearance of large-scale datasets (both public as well as commercial) that have provided researchers a wide view on crowds urban transportation and its inherent emerging patterns.
Once establishing the need for mobility modeling, as well as reviewing various approaching for it, we discuss ride sharing, and specifically -- its application as a component in the comprehensive urban design -- again, focusing on works that have taken a data-driven approach to its analysis.
We conclude this section by reviewing recent ride sharing optimization.

\subsection{Background --- Traffic Design as an Element of Modern Urban Design}

\subsubsection{Historical Overview}
The field of traffic design has traditionally been dominated by traffic engineers, advocating a distinct mechanistic character, in which the planning process was seen as a series of rigorous steps undertaken to measure likely impacts, mitigated by engineering solutions. The common practice revolves around a four-steps model:
(a) Trip generation; (b) Trip distribution; (c) Modal split, and (c) Route selection.
Each of those steps involves the use of various mathematical techniques, including base model regression, analysis of critical routes, entropy-maximizing models, etc.
The execution of this four steps model results in predictions regarding future traffic flows that are then used as input for assessment models, used to identify and evaluate planning options. Since the most common prediction is that present capacities will be unable to cope with traffic growth, the tendency has been to generate planning solutions that traditionally call for an expansion of capacity.
This approach has been referred to as ``Predict and Provide'', and has guided a major part of urban transport planning from the 1940s to the 1980s. One result of this dynamic has been an enormous expansion of highway construction that reinforced the dominance of the automobile in most large cities.

Traffic problems in cities have increased significantly since the 1970s, despite a great deal of urban transport planning. There is a growing realization that perhaps planning has failed and that the wrong questions have been asked. Rather than estimate traffic increases and then provide capacity to meet the expected growth, current designers speculate that a better management of the transport system might be achieved through new approaches to planning. Just as urban planning requires the inputs of many specialists, urban transport planning is beginning to utilize multi-disciplinary teams in order to broaden the scope of the planning process. Planning is still a multi-step process, but it has changed considerably.
Transportation planning can be thought of as a continuing, cooperative and multi-dimensional oriented problem-solving process, resulting in digested information provided to decision and policy makers. The process can be described in the following steps:

\begin{itemize}
  \item Goal setting (how do we see the “ideal future“?)
  \item	Problem identification (what are the problems in reaching our goals?)
  \item	Data collection and analysis (what is the current situation, projected issues, etc.?)
  \item	Alternative actions (what are possible solutions to coping with the problems?)
  \item	Evaluation of alternatives (what are the pros and cons of each solution?)
  \item	Recommended alternative (what is the best course of action to take?)
  \item	Implementation (carry out the solution or combination of solutions)
  \item	Monitor and feedback (what happened? did the solution(s) solve the problem? are there other problems?)
\end{itemize}

Whereas this process remained roughly unchanged throughout the years, there has been significant advancement in the various solutions employed.

\subsubsection{Mobility vs. Accessibility : the Conflict that Shapes Transportation Policies}

Transportation analysis is affected by how transport is defined and evaluated. As asserted in~\cite{justice1}, conventional planning tends to evaluate transport based on mobility (physical travel), using indicators such as traffic speed and roadway level-of-service. However, mobility is seldom an end in itself, the ultimate goal of most transport activity is accessibility, which refers to people’s ability to reach desired services and activities, and to do so in an affordable way, the latter being a subjective mixture of various economic, environmental, sociological and psychological considerations.
Various factors can affect affordable accessibility including direct and indirect monetary costs, transport network connectivity and ease of use, pollutants emission which at high levels may adversarially influence usability, the geographic distribution of urban activities, and mobility substitutes such as telecommunications and delivery services.

This has clear social implications, as mobility-based planning tends to favor faster travel modes at the expense of longer trips, over slower travel modes and shorter trips. Therefore, we can expect that alternatives that favor motorists will be preferred over alternatives for non-drivers. For example, evaluating transport system performance based on roadway level-of-service tends to justify roadway expansion projects, despite the tendency of wider roads and increased traffic speeds to degrade walking and cycling conditions.
As pointed out in \cite{justice1}, transportation planning often involves tradeoffs between social (or ``equity'') objectives and other planning objectives. For example, improving pedestrian safety may reduce traffic speeds (therefore, allegedly decreasing economic productivity), where providing public transit services may require tax subsidies, and in some cases may increase local air and noise pollution.

Viewing the design of urban transportation system through the prism of equality and affordable accessability, a dynamic self-optimizing ride-sharing system has a clear benefit, through its potential to significantly reduce various current inhibitors for car-sharing and ride-sharing solutions (such as overall travel times), thus facilitating its adoption by a more heterogenous crowd of urban travelers.

\subsubsection{Urban-Aware Transportation Design and the “Entrepreneurial City”}

In his seminal work \cite{justice2} Harvey had pointed out the ``entrepreneurial trend'' that cities in the last decades had undergone, featuring the refocus of municipal authorities attention for capital-generating oriented projects with a strong, and often central, participation of private investors and entrepreneurs. In his words:

\begin{quote}
The `managerial' approach so typical of the 1960s has steadily given way to initiatory and `entrepreneurial' forms of action in the 1970s and 1980s. In recent years in particular, there seems to be a general consensus emerging throughout the advanced capitalist world that positive benefits are to be had by cities taking an entrepreneurial stance to economic development. What is remarkable is that this consensus seems to hold across national boundaries and even across political parties and ideologies.
\end{quote}

As a result, it appears that the physical and social landscape of urbanization was significantly shaped in the last few decades according to aggressive bias towards capitalist and privatized criteria, constraining current and future non-capitalist and non-for-profit development thrusts.
It seems clear that this trend naturally blends with the appearance of various types of ride-sharing systems, as its inherent principle is maximizing  financial benefits (either by reducing the direct cost for the passengers \cite{li2015intelligent,nielsen2015white} or increasing the revenues for the systems' operators \cite{isaac2014disruptive,huefner2015sharing}) \cite{anderson2014not,morency2007ambivalence} \emph{through} the optimization of various non-capitalist values, such as reduced pollutant emission \cite{erdougan2015ridesharing}, aesthetic considerations (everybody uses the same type of car), better use of land areas (less parking spaces \cite{zhang2015exploring}), increased safety \cite{feeney2015ridesharing} (using well maintained public cars rather than poorly handled old models) and accessible private transportation for the elders~\cite{freund1997independent}.

\subsubsection{Environmentally-Aware Traffic Optimization and Monitoring}

Vehicle air pollutant emissions are considered one of the major environmental issues. The problem stems from a constant growth in motorization rate in relation to the technological means for reduction of pollutant emissions \cite{de2000air}.
Travel behavior, like many other aspects of daily life, is being transformed by information technology.
Accessibility can no longer be measured only in terms of travel time, distance or generalized travel cost.
Some of the potential effects of IT on transportation both personal and freight were explored in \cite{environmental-thomas2001}, and similar results were later demonstrated in \cite{environmental-larson2009}.
More general attempts to model human mobility patterns were discussed in \cite{gonzalez:2008nature}, \cite{song2010limits}, \cite{simini2012universal}, with further expansion in \cite{camara2012internet}, suggesting using the internet itself as a transportation oriented sensor.
In this respect, ride-sharing can serve an important role in reducing vehicle emission and improve air quality, with its main contribution being the reduction of the number of vehicles (through the increase in their utilization), alongside the improvement in their maintenance level.

\subsection{Recent Developments and the ``Transportation as a Service'' Concept}

The main evolution of private shared transportation schemes had gained momentum since the early 70s and is a popular research field until today, focusing mainly on the economic aspects of private shared transportation, examining among others demand and feasibility considerations.
However, in the recent decade there has been a significant shift in this research towards a data-driven analysis, using recently digitally available large-scale high-resolution datasets such as phone-call records or taxi-rides. In this sense, this is yet another step in the centaury-long journey for utilizing technological developments for the enhancements of urban transportation systems.

In year 2016 alone there has been an explosion in news about the future of transportation, mostly revolving around ridesharing and various technologies that will enable its high efficiency:

\begin{itemize}
  \item Ford announced plans for its own car-sharing service built around self-driving Fords \cite{lit_ford1}.
  \item Elon Musk penned a second master plan envisioning a future car-sharing service built around self-driving Teslas \cite{lit_musk1}.
  \item Nutonomy launched a trial in Singapore of its own ride-sharing service built around Renault and Mitsubishi vehicles modified to be self-driving \cite{lit_Nutonomy1}.
  \item Uber announced its own self-driving trial in Pittsburgh in partnership with Volvo \cite{lit_Uber1}. Uber also acquired self-driving startup Otto, founded by former members of Google’s self-driving team \cite{lit_Uber2}.
  \item Alphabet (Google's holding company) announced an expansion of its Waze-based ride-sharing service from Israel to Uber’s home city of San Francisco \cite{lit_Alphabet1}.
  \item In August 1st 2016 the Chinese ridesharing giant Didi acquired Uber China, creating a company that serves 300 million users using 1.8 million drivers, estimated in 35 Billion US valuation \cite{lit_didi1}.
  \item Another Asian ridesharing service, called \emph{GrabTaxi}, active in 6 countries in South-east Asia, announced the expansions of its R\&D center to the US \cite{lit_grab1}.
\end{itemize}

These recent developments are all aiming towards the common goals of \emph{transportation-as-a-service}, as stated by Uber CEO Travis Kalanick \cite{lit_Kalanick1}:

\begin{quote}
The minute it was clear to us that our friends in Mountain View\footnote{Google Inc} were going to be getting in the ride-sharing space, we needed to make sure there is an alternative\footnote{A self-driving car}. Because if there is not, we’re not going to have any business. Developing an autonomous vehicle is basically existential for us.
\end{quote}

As analyzed by Ben Thompson \cite{lit_Thompson1}, this process takes place in 4 steps:

\subsubsection{Transportation as a Service 1.0: UberX : }
In this stage of the technological evolution, drivers and riders are the most important components, as drivers bring their own cars, existing mapping solutions are good enough, and routing is relatively simple. Managing millions of matches a day between drivers and riders is a complicated undertaking, and the availability of the underlying algorithms that enable it becomes ever more critical.

\subsubsection{Transportation as a Service 1.5: UberPool : }
The UberPool service that enables passengers to share a ride introduced several changes to the ridesharing system: whereas the drivers, cars and mapping components remained roughly unchanged, the routing requirement underwent a massive increase in complexity over UberX. In addition, there has also been a change in passengers' expectations (and behavior with regard to pick-up and drop-off points) relative to UberX, allowing in turn for routing solutions that rely on a longer wait time (in order to enable an improvement in the ridesharing utilization \cite{angelelli2016proactive,stiglic2016making}).
Uber investor Bill Gurley recently commented that getting the algorithms behind UberPool right is an incredibly complex problem, and the only real way to solve it is to slowly but surely work out heuristics that work in real world situations \cite{lit_Bill1}. This further highlights the importance of predictive methods with respect to ridesharing optimization.

\subsubsection{Transportation as a Service 2.0: Human Self-Driving Cars : }
Recently, Google has launched its own ridesharing service, through its Waze service, capping the price at 54 cents a mile, which is the IRS standard mileage rate\cite{lit_irs1}. This was designed in order to prevent the possibility of earning money by simply driving-for-hire. The claimed intent behind this move was to facilitate the sharing of gas money (and carbon footprint) while eliminating the various contractual and legal aspects of previous ridesharing services.
This marks an important shift in the components in transportation-as-a-service, as drivers can now be considered as effectively free, since they are already headed towards the desired destination. In addition, the cars they are using have already been bought for the purpose of transporting the driver, so they can be considered as effectively free as well.

\subsubsection{Transportation as a Service 3.0:  Self-Driving Cars : }
Whenever self-driving cars do arrive, nearly every piece in the transportation-as-a-service puzzle will be significantly affected:
Drivers will, by definition, not exist. Cars would be self-driving, which means they will likely be very expensive at least at the beginning, requiring significant capital costs. Having large fleets of fully-autonomous cars would also significantly increase the level of details the mapping systems would have to possess, while still maintaining the (very high) complexity level of UberPool -- de-facto further increasing the overall complexity\cite{cordeau2007vehicle}. It is therefore likely to assume that such system would have to rely, at least to a certain extent, on various predictive technologies, that would provide lookahead capabilities to their operational heuristics \cite{savelsbergh1998drive}.

\subsection{Data Driven Mobility Modeling}

Data analysis offers ways of identifying and predicting interesting events or extracting information in a variety of contexts.
The capacity to collect and analyze massive amounts of data has transformed many scientific fields. One of the
most recent disciplines to benefit from the power of data is network science~\cite{Lazer06022009,SBP-Trends}, investigating data
at large scale to reveal patterns of individual and group behaviors.
Network features can signal and are often used to predict events or properties that are external to the network, but influence it.
A network can often be built on easily available data and serve as an important source for predictions regarding various (seemingly unrelated) events and large-scale decision-making processes \cite{altshuler2015social,krafft2016human,Altshuler-trends-arxiv-2011}.
Features of a phone call network can signal the occurrence of an emergency situation~\cite{altshuler2013social}
or predict trust among individuals~\cite{6804686}, and specific behaviors in a Twitter account can identify a spammer~\cite{Almaatouq:2014:TGC:2615569.2615688}.

Such discoveries had in turn sparked the interest of researchers in different research fields, who could benefit from this new ability to model large-scale human dynamics. One of the fields most influences by this evolving research thrust was the data-driven study of human mobility and its potential application for Intelligent Transportation Systems \cite{gonzalez:2008nature,altshulerrationality,environmental-armstrong,alexander2015assessing,puzis2012augmented}.

It has been recently shown that in trying to detect semantic network events (such as an accident or a traffic jam) it is crucial to understand the underlying structure of the network these events are taking place at \cite{waclaw2007statistical,wassermann1994social,altshuler2013detecting}, the role of the link weights \cite{granovetter1973strength}, as well as the response of the network to node and link removal \cite{albert2000error}. Past research \cite{onnela2007structure} had pointed out the existence of powerful patterns in the placement of links, or that clusters of strongly tied together individuals tend to be connected by weak ties \cite{granovetter1973strength}. It was also shown that this finding provides insight into the robustness of the network to particular patterns of link and node removal, as well as into the spreading processes that take place in the network \cite{pastor2001epidemic,pastor2007evolution}. In addition, recent work had demonstrated the trade-off between the number of individuals (the width of the data) and the amount of information available from each one (the depth of the data), with respect to the ability to accurately model  crowds behavior \cite{SBP-Crowd,ARXIV-learning-with-time,altshuler2013trade,altshuler2014campaign,altshuler2015campaign}. An analytical approach to this problem discussing the (surprisingly large) amount of personal information that can be deduced by an ``attacker'' who has access to one's personal interactions' meta-data can be found in \cite{SR-IIS,gaydar-jernigan,pan2012decoding}.

One of the first works that examined the statistical distribution of event appearance in mobility and communication networks have found that these follow a power law principle \cite{barabasi2005origin}, and that such distribution is significantly affected by anomalous events that are external to the networks \cite{candia2008uncovering}.
A method for filtering mobile phones Call Data Records (CDRs) in space and time using an agglomerative clustering algorithm in order to reconstruct the origin-destination urban travel patterns was recently suggested in \cite{alexander2015origin}.
Other works had examined the evolution of nodes groups in the network, aiming for the development of algorithms capable of identifying ``new clusters'' -- a certain kind of anomalous network pattern~\cite{palla2007quantifying}.

Recent works that have been analyzing data collected by the pervasive use of mobile phones have broadly supported the notion that most of human mobility patterns are affected by a relatively small number of factors, easily modeled, and very predictable \cite{brockmann2006scaling,de2013unique,song2010limits,calabrese2013understanding}
A comprehensive survey of ride-sharing literature can also be found in \cite{furuhata2013ridesharing} and another recent relevant study
that developed spatial, temporal, and hierarchical decomposition solution strategy for ride-sharing is presented in \cite{ghoseiri2012dynamic}.
Another study was able to show that using these models, the analysis of a large number of phone calls is able to show that the current statutory boundaries between provinces in Belgium are erroneous, in the sense that they inaccurate partition the country to cultural and economic regions, as those can be well defined through the analysis of phone calls, under the assumption that people mainly call their family and friends
\cite{blondel2010regions,blondel2008fast}.

A recent study \cite{jiang2016timegeo} tackled the problem of mobility modeling when a highly sparse dataset is available.
The authors presented a mechanistic modeling framework (TimeGeo)
that can generate urban mobility patterns with resolution of 10 minutes and hundreds of meters. It ties together the inference of
home and work activity locations from data, with the modeling of flexible activities in space and time. The temporal choices
are captured by only three features: the weekly home-based tour number, the dwell rate, and the burst rate. These combined generate
for each individual: (i) stay duration of activities, (ii) number of visited locations per day, and (iii) daily mobility networks. These parameters
capture how an individual deviates from the circadian rhythm of the population, and generate the wide spectrum of empirically
observed mobility behaviors.

\subsection{Ride Sharing -- a Data Driven Analytic Approach}

To-date, much of the research related to ride sharing has focused on understanding the characteristics of ride sharing
trips and users. In a recent survey of app-based, on-demand rideshare users in San Francisco, researchers found that 45\%
of ridesharers stated they would have used a taxi or driven their own car had ridesharing not been available, while 43\%
would have taken transit, walked, or cycled \cite{rayle2014app}.

A recent work by Santi~et~al.~\cite{santi2014quantifying} introduces a way of quantifying the benefits of sharing. The study applies to a GPS dataset of taxi rides in New York City and uses the notion of \textit{shareability network} to quantify the impact and the feasibility of taxi-sharing.
When passengers have a 5 minutes flexibility on the arrival time, and they are willing to wait up to 1 minute after calling the cab, over 90\% of the sharing opportunities can be exploited and 32\% of travel time can be saved.
The authors have also shown that the problem is computationally tractable when we look for sharing a taxi among two people with the option of in-route picking up.
Furthermore, sharing solutions involving more people are not tractable, but do not provide a significant improvement with respect to solutions involving only two people.
Similar results have been demonstrated using a theoretical model analyzing \emph{Autonomous Mobility On Demand} system, demonstrating that a combined predictive positioning and ridesharing approach is capable of reducing customer service times by up to 29\% \cite{miller2016predictive}.

An extensive simulation infrastructure for ride sharing analysis is suggested in \cite{ota2016stars}, allowing the initialization and tracking of a wide variety of realistic scenarios, monitoring the performance of the ride sharing system from different angles, considering different stakeholders’ interests and constraints. The simulative infrastructure is claimed to use an optimization algorithm that is linear in the number of trips and makes use of an efficient and fully parallelized indexing scheme.

In another paper \cite{adler2016optimal} a system in which vehicles arrive at a station according to a stochastic process was discussed, such that the vehicles may wait for each other in order to form platoons to save energy, but at the cost of incurring transportation
delays. The authors explored this trade-of between energy consumption and transportation delays, finding a Pareto-optimal boundary and characterization of the optimal polices in both the open-loop and feedback regimes.

In another study by Cici~et~al.~\cite{cici2014assessing} mobile phone data and social network data were used to estimate the benefits of ride sharing on the daily home-work commute.
Mobile phone data are easier to collect than GPS traces, and have a higher penetration, providing a good sample of a city mobility.
Social network data are used to study the effect of friendship on the potential of ride sharing, showing that if people want to travel only with friends then expected ride sharing benefits are negligible. On the other hand, when people are willing to ride with friends of friends the achieved efficiency resembles this of the variant that also allow riding with strangers (implying that safety issues may have significant effect on the actual success of a ride sharing solution).

Similar study have been presented by \cite{tachet2016scaling} calculating shareability curves using millions of taxi trips in New York City, San Francisco, Singapore, and Vienna, showing that a natural rescaling collapses them onto a single, universal curve. The authors presented a model that predicts the potential for ride sharing in any city, using a few basic urban quantities and no adjustable parameters.

A related study \cite{stiglic2016enhancing} analyzed the assumption that seamless integration of ride-sharing and public transit may offer fast, reliable, and affordable transfer to and from transit stations in suburban areas thereby enhancing mobility
of residents.

The growing use and popularity of smart phones and GPS-enabled devices provides tools that facilitate the efficient implementation of ride sharing. However, privacy and safety concerns are key obstacles faced when encouraging people to use such services. In the work of \cite{goel2016privacy} the authors have presented \emph{``Match Maker''}, a negotiation-based model that hides exact location information data for system participants while implementing privacy preserving ride sharing. The algorithm is based on the concept of imprecision (not being precise about location of the user out of set of n locations) and follow the idea of obfuscation, which equates a higher degree of imprecision with a higher degree of privacy. The work discusses two attack types that could circumvent privacy preserving ride sharing, comparing their suggested algorithm with the standard central trusted server model collecting precise location data. The authors show that while preserving passengers' privacy ride sharing can still be implemented in a way that saves between 9\% and 21\% of vehicle KMs if drivers are only prepared to accept slight detours of their usual trips, using data originated in the city of Melbourne.

The issue of pricing policies in ride sharing services have gained significant attention recently. with the booming expansion of commercial ride sharing services such as \emph{Uber}, \emph{Lyft} and others. The work of \cite{banerjee2016dynamic}
studies dynamic pricing policies for ride sharing platforms. As such platforms are two-sided this requires economic models that capture the incentives of both drivers and passengers. In addition, such platforms support high temporal-resolution for data collection and pricing. The combination of the latter requires stochastic models that capture the dynamics of drivers and passengers in the system.

In \cite{bimpikis2016spatial} the authors highlight the impact of the demand pattern of the underlying network on the platform’s optimal profits and aggregate consumer surplus. In particular, the authors establish that both profits and consumer surplus are maximized when the demand pattern is balanced across the network’s locations. In addition, the authors show that profits and consumer surplus are monotonic with the ``balancedness'' of the demand pattern (as formalized by the pattern’s structural properties). Furthermore, the widely adopted compensation scheme that allocates a constant fraction of the fare to drivers is explored, identifying a class of networks for which it can implement the optimal equilibrium outcome. The authors however showcase that generally this scheme leads to significantly lower profits for the platform than the optimal pricing policy especially in the presence of heterogeneity among the demand patterns in different locations.

The work of \cite{dai2016ridesharing} proposes a recommendation framework to predict and recommend whether and where should ride sharing users wait in order to maximize their chances of getting a ride. In the framework, a large-scale GPS data set generated by over 7,000 taxis in a period of one month in Nanjing, China was autilized to model the arrival patterns of occupied taxis from different sources.
The underlying road network was first grouped into a number of road clusters. GPS data were categorized to different clusters according to where their sources were located. A kernel density estimation approach was then used to personalize the arrival pattern of taxis departing from each cluster rather than a universal distribution for all clusters. Given a ``query'', the potential of ride sharing and where should the user wait by investigating the probabilities of possible destinations based on ridesharing requirements can then be computed.

\begin{figure}[htbp]
\centering
\includegraphics[bb=0 0 1000 800, clip=true, scale =0.3]{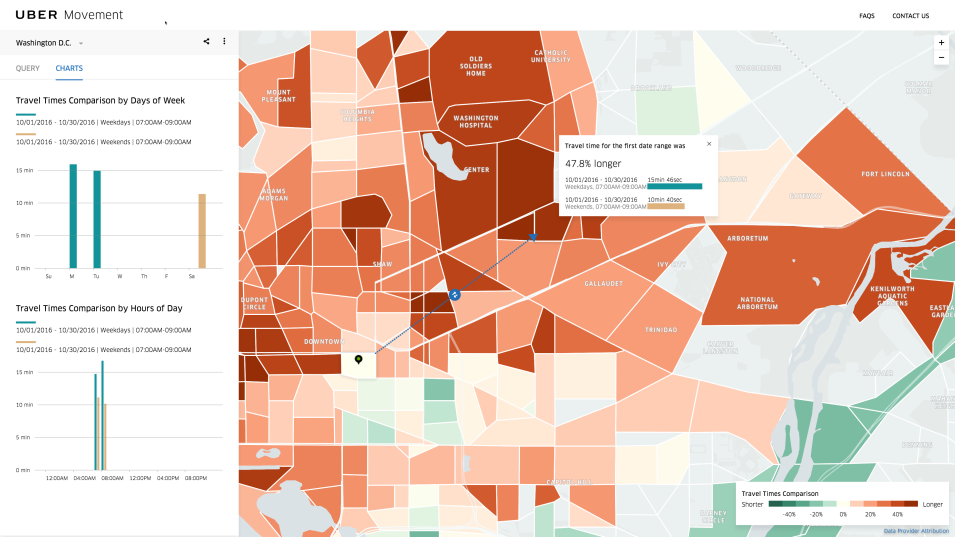}
\caption{Screenshot of \emph{Uber Movement}}
\label{fig:ubermovement1}
\end{figure}

\begin{figure}[htbp]
\centering
\includegraphics[bb=0 0 1000 800, clip=true, scale =0.3]{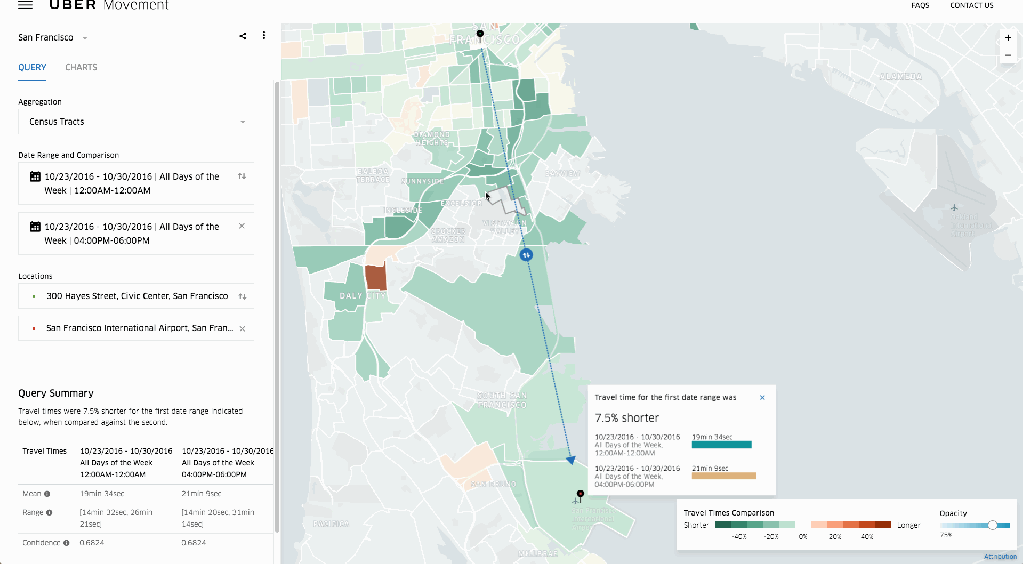}
\caption{Screenshot of \emph{Uber Movement}}
\label{fig:ubermovement3}
\end{figure}

Recently, \emph{Uber}, one of the world's largest ridesharing operators, announced that it is making its dataset available to researchers worldwide. The ride-hailing company’s new \emph{Movement} website \cite{Uber_Movement} will offer up access to its data around traffic flow in scores where it operates, intended for use by city planners and researchers looking into ways to improve urban mobility.
As the company possess a highly rich data containing insights into how traffic works within a city, and it can anonymize this data so that it isn’t tied to specific individuals in most cases. According to the company, this data could be used to address problems city officials and urban planners encounter when they’re forced to make key, transformational infrastructure decisions without access to all of, or the proper information about actual conditions and causes.
Essentially, according to Uber, it’s hoping to make it easier for those with influence over a city’s transportation picture to make the right decision, and to be able to explain why, where and when the changes are happening with accurate data backing them up. It also wants to do this in a way that makes it easy for organizations to work with, so it’s releasing the data organized around traffic analysis zones within cities, which are agreed-upon geographic demarcations that help with existing urban planning and traffic management.
Users of the website can control parameters such as time of day, day of week and zones to call up Uber’s data for that specific point or range, and can download the data, both with existing time series charts and in raw format for inputting into their own models (see illustrations in Figures \ref{fig:ubermovement1} and \ref{fig:ubermovement3}).

\subsection{Existing Ride Sharing Optimization Models}

\begin{figure}[htbp]
\centering
\includegraphics[bb=0 0 500 300, clip=true, scale =0.4]{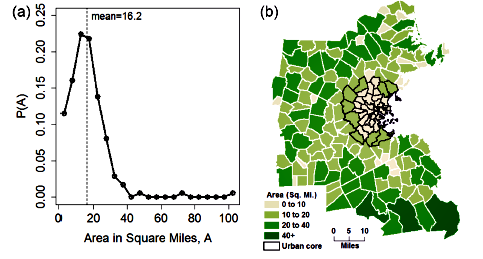}
\caption[Probability and spatial distributions of community area in miles$^{2}$]{(a) Probability distribution of community area in miles$^{2}$ and (b) Spatial distribution of community area, with area increasing from light to dark green, and black borders denoting the Urban core.
The figure is based on the result presented in \cite{alexander2015assessing}.}
\label{fig:alexander1}
\end{figure}

The recent work of Alexander and Gonzalez \cite{alexander2015assessing} uses smart-phone data in order to model the behavior of an urban population in Boston, in an attempt to assess the impact of efficient ridesharing service on the urban traffic, and specifically on the expected levels of congestion.
As can be see in Figure \ref{fig:alexander1} that is taken from that work, this data-centric approach leads to a highly accurate modeling of the mobility patterns in the city. However, much like most of the recent work on this subject, the researchers have followed an aggregative modeling, that tries to find the static long-term definitive mobility patterns, purposely omitting any dynamic fluctuations.

In another study, researchers from the Microsoft Research Center \cite{where-to-find-my-next-passenger} analyzed the ride data of 12,000 taxis during 110 days in order to model the mobility patterns of potential passengers. Using this probabilistic model the researchers were able to build a recommendation system for taxi drivers that would maximize their profits (yielding an overall 10\% improvement in the overall profits) and a second recommendation system for passengers, advising them where to turn in order to maximize their chances of finding a vacant taxi (with 67\% accuracy).
A similar research can be found in \cite{li2012prediction}.

A recent review of dynamic ridesharing systems \cite{agatz2012optimization} focused on the optimization problem of finding efficient matches between passengers and drivers. This ride-matching optimization problem determines vehicle routes and the assignment of passengers to vehicles considering the conflicting objectives of maximizing the number of serviced passengers, minimizing the operating cost, and minimizing passenger inconvenience.
Another study \cite{wang2016activity} presented an algorithm that increases the potential destination choice for ride sharing schemes set by considering alternative destinations that are within given space-time budgets.

A comprehensive survey of car-sharing related works that examined a variety of techniques designed to improve the efficiency of ride-matching schemes contains no indication of the use of predictive techniques in order to gain such improvement, further highlighting the emphasis of past works mainly on the optimization itself, rather on the development of ways to foresee it \cite{furuhata2013ridesharing}.

A recent work modeled the mobility patterns of 5 million residents using data from ``smart card'' used for the metropolitan buses and rail system.
This vividly demonstrates that an individual’s encounter
capability is rooted in his/her daily behavioral regularity \cite{sun2013understanding}. This explains the emergence of ``familiar strangers'' in daily life -- patterns of synchronicity that emerge spontaneously. Strikingly, it is shown that individuals with repeated encounters are not
grouped into small communities, but become strongly connected over time, resulting in a large, but imperceptible, small-world contact
network or ``structure of co-presence'' across the whole metropolitan area. Such structures, when detected, can in turn be used in order to further improve ride-sharing utilization, as they bear predictive information regarding future synchronization -- in turn, a strong prediction of high ridesharing utilization.

\remove{
\begin{figure}[htbp]
\centering
\includegraphics[bb=700 0 1300 700, clip=true, scale =0.3]{sun2013_1.png}
\caption[Illustration of ``familiar strangers'']{Illustration of ``familiar strangers'' -- passengers who shared the same origin and destination stations. The size of the circles indicates the size of the groups. The figure was published in \cite{sun2013understanding}.}
\label{fig:sun2013_1}
\end{figure}

\begin{figure}[htbp]
\centering
\includegraphics[bb=0 0 1300 620, clip=true, scale =0.25]{sun2013_2.png}
\caption[Probability density function of trip frequency and number of encountered people, and interevent interval between consecutive bus trips]{(C) Illustrates the probability density function of trip frequency and number of encountered people of all individuals. (D) Probability density function of trip frequency and interevent interval between consecutive bus trips. Although diversity of
individual transit use obscures the distribution, a remarkable area with ($5 < f < 25 ; 0 < \tau < 25$) can be identified through similarity in regular commuters’ transit behavior.
The figure was published in \cite{sun2013understanding}.}
\label{fig:sun2013_2}
\end{figure}
}

On a similar note, a recent study \cite{stiglic2015benefits} analyzed the benefits of meeting points in ride-sharing systems, investigating the potential benefits of introducing meeting points in a ride-sharing
system. With meeting points, riders can be picked up and dropped off either at their origin
and destination or at a meeting point that is within a certain distance from their origin or
destination. The increased flexibility results in additional feasible matches between drivers
and riders, and allows a driver to be matched with multiple riders without increasing the
number of stops the driver needs to make. A similar approach for the optimization of such meetings points was discussed in \cite{goel2016optimal}.

The challenge of rides-matching was also discussed in works such as \cite{alarabi2016demonstration}, \cite{stiglic2016making} or \cite{alonso2017demand}, which have demonstrated that 2,000 vehicles (15\% of the taxi fleet in New York) of capacity 10 passengers (or 3,000 vehicles of capacity of 4 passengers) can serve 98\% of the New York taxi demand within a mean waiting time of 2.8 min and mean trip delay of 3.5 min.

A \emph{path merging} approach, which instead of merging rides to and from the same locations calculate new paths which goes through the same locations of the original trips, at the same order, and thus improves the ability to merge rides, was discussed in \cite{d2016path}.

In a recent theoretical study \cite{wu2016clustering} where the combinatorial optimization of ridesharing matching problem was tackled using the proof of the equivalence between classical centroid clustering problems and a special case of set partitioning called metric k-set partitioning, in
which an efficient expectation maximization algorithm was used to achieve a 69\% reduction in total vehicle distance, as compared with no ridesharing.

A fully decentralized reputation-based approach is discussed in \cite{sanchez2016co}, using a peer-to-peer architecture to provide self-assembling ride sharing infrastructure capable of functioning with no central authority or regulator.

\section{Dataset and Methodology}
\label{sec:data}

Our analysis was performed using a dataset of 14,776,615 taxi rides collected in New York City over a period of one month (January 2013)~\cite{nyc}.
Each ride record consists of the following fields: pick-up time, pick-up longitude, pick-up latitude, drop-off longitude, drop-off latitude, number of passengers per ride, average velocity and overall trip duration. Times granularity is second-based and positional information has been
collected via GPS technology by the data provider.
From this raw data sample we omit records containing missing or erroneous GPS coordinates, as well as records that represent rides that started or ended outside Manhattan, yielded a cleaned dataset containing 12,784,243 rides.

As a first step in modeling the feasibility and efficiency of ride-sharing schemes using taxi rides in New York City, a comprehensive understanding of the data itself is required. How do the rides distribute over the various geographic locations? Are there patterns that emerge when observing the O-D matrix of the various rides? Can we use those in order to predict the destinations of passengers when they board a taxi at a certain location?
The figures below attempt to answer some of the Power Low distribution) strongly implies on the potential of a network-centric approach as the method of choice with respect to the modeling of the dynamics of the data.

Some of the following illustrations analyzing the dataset's statistical properties were first presented in our previous publication \cite{ICCSS15Altshuler} as well as by \cite{shmueli2015ride} in their analysis of this public dataset. These illustrations appear here to contribute to the reader's understanding of the nature of the data and the behavior dynamics it encapsulates.

\begin{figure}[h!]
\centering
\includegraphics[bb=0 0 600 500, clip=true, scale =0.3]{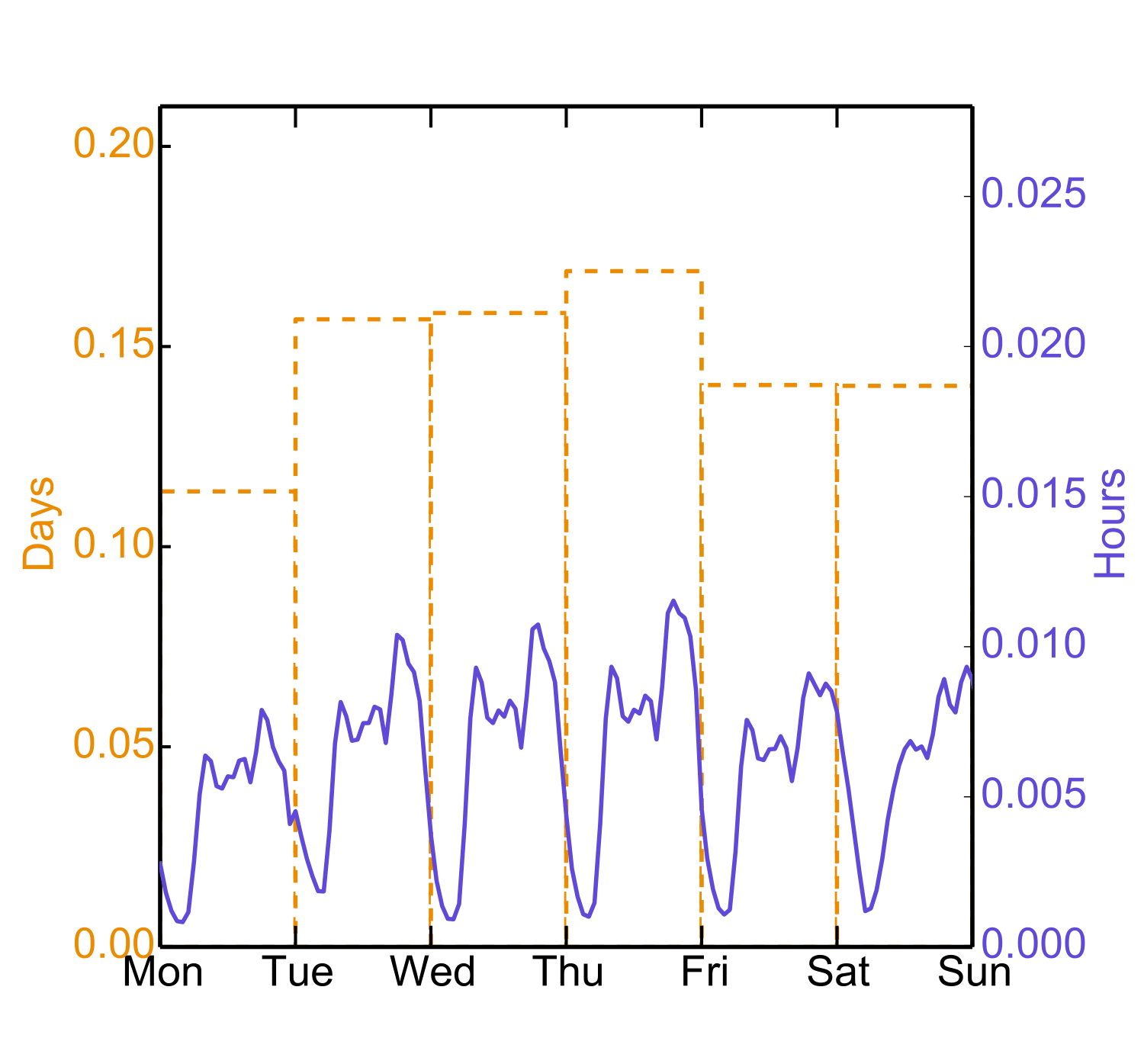}
\caption[Probability Density Function of the number of rides per day of week/hour of day]{Probability Density Function (PDF) of the number of rides per day of week/hour of day. Afternoon peaks are centered on average around 7pm (Chart was taken from \cite{shmueli2015ride}).}
\label{fig:pdf_time}
\end{figure}

Figure~\ref{fig:pdf_time} reports the distribution of rides per day of the week and per hour of the day.
As can be seen in the figure, the number of rides has a far-from-uniform time distribution.
More specifically, the number of rides is higher in the middle of the week and is lower during the weekend. In addition, the daily rides distribution peaks, as expected, in the morning hours and around 6-7pm.

We use the set of taxi ride records to construct a ``rides network'' $G_{T_{1},T_{2}}$, comprising of $|V|$ nodes representing equally sized squared regions of New York City, and a set of $|E|$ edges, such that each edge $(u, v) \in E$ corresponds to a connection between two regions $u, v \in V$ if and only if there exists at least one ride from region $u$ to region $v$ in the time-frame referred to by the network. Such a connection exists if and only if a ride started at some time $t$ departing at $u$ and reaching $v$, or vice versa, such that $T_{1} \leq t \leq T_{2}$ is contained in the time period defined for the network $G_{T_{1},T_{2}}$.

As we create edges only based on rides that were created during a certain period of time the network may change (and quite significantly so) for various values selected for $T_{1}$ and $T_{2}$. As the time period defined by these values increases the network is expected to contain more edges, with the densest network received for $G = G_{-\infty,\infty}$ being the network that is based on the complete aggregation of all the rides. In order to encapsulate the traffic properties of a certain point in time $T$ we would observe the time period circumventing $T$. Similarly, in order to analyze the network dynamics, that is -- the way it changes over time, we would analyze the evolution of the network properties for networks created in non-identical yet partially-overlapping time periods. This methodology is extensively used in Section~\ref{sec:network}.

\begin{figure}[h!]
\centering
\includegraphics[bb=0 0 400 300, clip=true, scale =0.5]{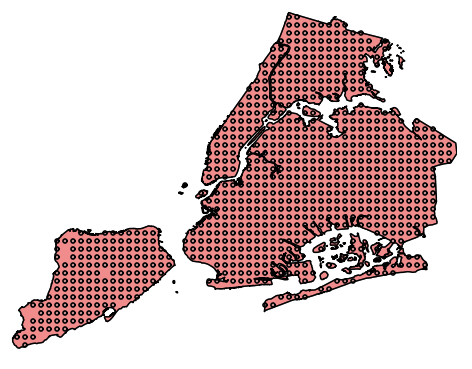}
\caption[Illustration of the rides network $G$, portrayed on the map of NYC]{Illustration of the rides network $G$, portrayed on the map of NYC. It can be seen that the network has high density through the city, with a few empty spots in Staten Island. The figure is based on the result presented in \cite{shmueli2015ride}.}
\label{fig:nodesNYC}
\end{figure}

For different granularity of city partitioning (reflecting through the use of different sizes of the square regions) different ride networks would be produced. However, Network Theory implies that changing this parameter would not affect the existence of various mathematic invariants such as the network's ``Scale Free-ness'' or its expected small diameter \cite{CSS-BarabasiAlbert-Science-1999}, but rather -- mainly change the sparsity of the network and its number of nodes.
During this work we have examined several sizes of squared-regions, ranging from rectangular regions of 0.0156 square miles in size, to 1 square mile, obtaining similar results. The analysis below is based on square tiles of 0.39 square mile (i.e. 1 square kilometer). In such a case, when taking $G = G_{-\infty,\infty}$, the network that aggregates all the rides, it comprises of 813 nodes and 58,014 edges.
Figure~\ref{fig:nodesNYC} illustrates the geographical distribution of the nodes $V_{G}$ on the map of New York.

\begin{figure}[htbp]
\centering
\includegraphics[bb=0 0 400 300, clip=true, scale=0.5]{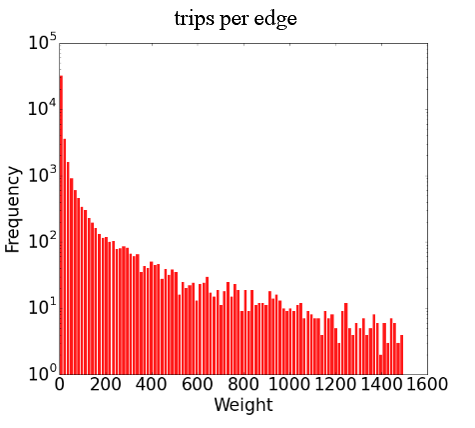}
\caption[Edge weights of the taxi rides network $G$]{Edge weights of the taxi rides network $G$, denoting the number of trips per edge (namely, between every two nodes in the city). The X-axis denotes the number of trips per edge (representing a pair of origin-destination nodes), and the Y-axis (shown in a log-scale) represents the number of edges who have such number of trips.}
\label{fig:stat1}
\end{figure}

Figure~\ref{fig:stat1} illustrates the distribution of the number of trips on the various O-D routes in the taxi network. By \emph{weight} we refer to the number of trips that took place through this edge and by \emph{Frequency} we refer to the number of edges that have a specific weight. Note the small number of edges who have more than 500 rides (approximately 5,000 edges out of 58,000 edges). Similarly, over 47,000 edges has less than 50 rides passing through them. This observation coincides well with the fact that human mobility is known to follow a power low distribution \cite{gonzalez2008understanding}.

As we analyze the network properties of graph implied by the taxi rides, it is interesting to observe the characteristics of the degrees of the nodes of the network $G$. A `degree' of a node $v \in V$ is the number of nodes $v$ is connected to through edges in $E$, where such nodes represent the actual destinations passengers who boarded a taxi at location $v$ chose to go to. Namely, a degree of a node $v$ represents therefore the number of possible destinations a passenger boarding a taxi on location $v$ may chose to go to.
An important observation is that the popularity of a node $v$ as reflected both by its in-degree (i.e. the number of origins passengers depart from in order to get to $v$) as well as by its out-degree (i.e. the number of destinations passengers leaving $v$ may go to) is independent of the geographic size or shape of node $v$ -- as all nodes refer to equally-sized square regions.

Interestingly, analyzing the distribution of this property reveals that whereas there are some nodes with a high degree (probably corresponding to main train stations or large administration facilities) the vast majority of the nodes have a very low degree. In other words -- for the vast majority of the locations in New York, it is extremely easy to predict the destination of a passenger starting his ride there (as a low degree implies a low number of possible destinations, and a high chance of guessing the correct one).
This observation is quite remarkable, as it implies that taxi users are \emph{much more predictable} than may seem. Indeed, it seems that when one boards a taxi one's destination can quite accurately be predicted.

\begin{figure}[htbp]
\centering
\includegraphics[bb=-100 0 1200 300, clip=true, scale=0.4]{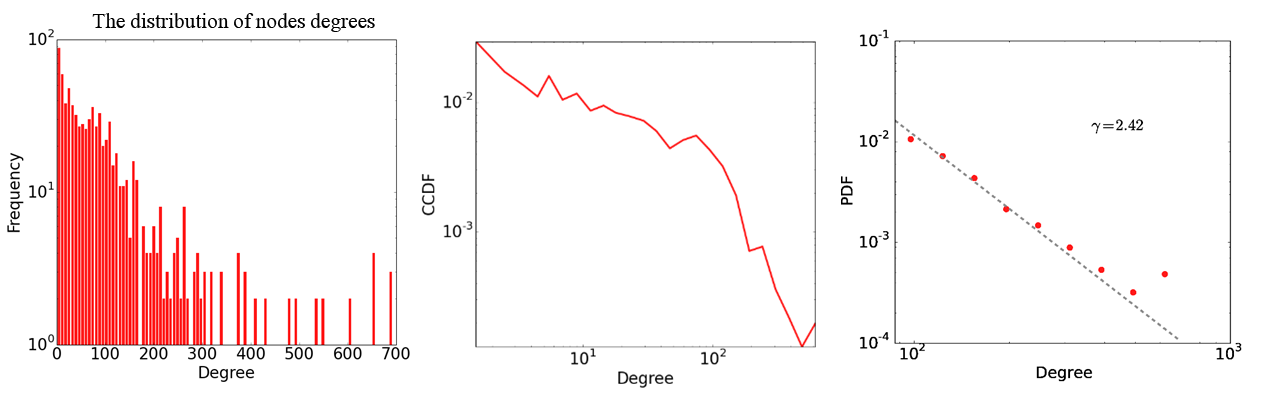}
\caption[The distribution of nodes degrees in the taxi rides network]{The distribution of nodes degrees in the taxi rides network, representing the number of possible destinations a passenger boarding a taxi at some location in the city may chose to go to. Note the surprisingly high number of origins with very low degrees -- number of possible destinations. The nodes' degrees are unaffected by the size or shape of the actual geographic region they refer to as all nodes refer to equally-sized square patches of the city.}
\label{fig:stat2}
\end{figure}

Specifically, in 24\% of the possible origins of a taxi ride in New York City, the number of possible destination of a passenger leaving these origins is on average 5, and in 43\% of the origins it is 10. A quick arithmetics yields that if at some point in time we would pick a random person just boarding a taxi anywhere in New York, we would have more than 7.5\% of guessing precisely his or her destination. This probability is about three times \emph{higher} than rolling a ``Snake Eyes'' (two 1's in a 6-sides dice).
See Figure~\ref{fig:stat2} for more details.

In this context it is also important to note that in this work we are less interested in the specific characterization of nodes having high (or low) degrees, but rather -- in the dynamics those values represent over time, as discussed in detail in the following sections.

In order to analyze the `sharability', or the ability to merge rides using the same vehicle at an overlapping times, we applied a simplified version of the methodology used by Santi et al.~\cite{santi2014quantifying} to calculate the potential benefits of ride sharing:
Let $T_i = (o_i,d_i,t^o_i,t^d_i),i=1...k$ be $k$ trips where $o_i$ denotes the origin of the trip, $d_i$ the destination, and $t^o_i,t^d_i$ the starting and ending times, respectively. We say that multiple trips $T_i$ are shareable if there exists a route connecting all of their origins and destinations in any order where each $o_i$ precedes the corresponding $d_i$.

Sharability, or `ridesharing utilization' is expressed in terms of the number of rides that can be `merged', as a function of the guaranteed quality of service, expressed through the number of latency minutes agreeable by the passengers --- the maximum time delay in catching a ride and arriving at destination, representing the maximum discomfort that a passenger can experience using the service. In other words, given a pre-defined level of discomfort passengers are willing to undertake (expressed in a prolonged wait-time), the ridesharing utilization depicts the portion of rides that are redundant and can be saved by merging with other rides to and from the same locations.

Our analysis aims at finding pairs of rides, which are represented in the network by the same edge (i.e., have the same origin and destination), that can be shared.
For each edge, we examine its corresponding set of originating rides, and count the number of ride pairs that can be merged, taking into consideration the maximum time delay parameter.

The main difference between our approach and the one discussed in \cite{santi2014quantifying} is that we only merge rides that leave the same origin `tile' and go to the same destination `tile'. There are several advantages for this approach:
\begin{enumerate}
  \item The routing-agnostic scheme is significantly less sensitive the temporary changes in the infrastructure, such as detours, traffic jams, accidents, and so on.
  \item Merging rides based only on their origin and destination makes our ridesharing policy entirely agnostic to the routing decision of the driver. Alternatively, the approach that is based on allowing rides to be merged even if they do not leave from the same origin, but are rather partially overlapping, depends on the assumption that the route of the ``containing ride'' indeed passes through the origin of the second ride. This assumption in turn depends on either perfectly guessing the routing decisions of the driver, or -- dictating those decisions to the driver by the ridesharing service.
  \item As a result, our routing-agnostic approach is also expected to be easier to implement in real-life scenario, as it requires less cooperation from the drivers.
  \item In addition, the increased simplicity of the routing-agnostic approach makes it easier to optimize from a computational point of view. The routing-aware approach discussed in \cite{santi2014quantifying} has a time complexity of $O(n^{2} \log(n))$ when merging pairs of rides \cite{galil1986efficient}, becomes much harder when triple rides merging is allowed \cite{chandra2001greedy}, and eventually becomes computationally unfeasible for larger numbers of rides-to-be-merged \cite{santi2014quantifying}.
  \item When comparing the merging efficiencies of our proposed routing-agnostic approach with the routing-aware one it is shown that whereas the latter is slightly more efficient when long wait-times are allowed (increasing our proposed 73\% sharability to 93\% for 5 minutes maximal delay), the improvement for shorter wait times becomes significantly smaller (this is illustrated by comparing Figure \ref{fig:monthly_benefit} below to Figure 3 in \cite{santi2014quantifying}).
\end{enumerate}

\begin{figure}[h!]
\centering
\includegraphics[bb=0 0 1500 1200, clip=true, scale =0.15]{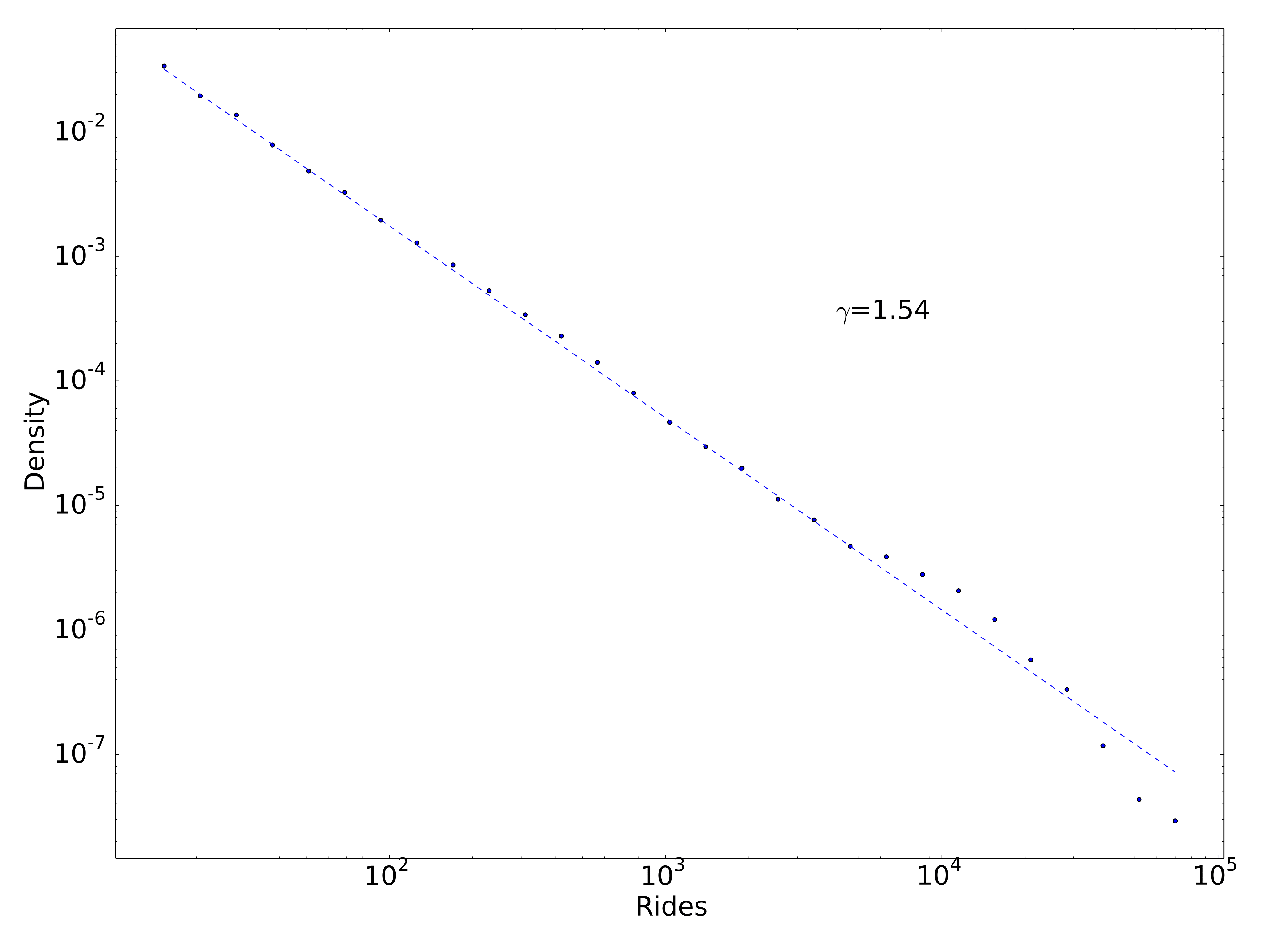}
\caption[Probability Density Function of the number of rides per edge]{Probability Density Function of the number of rides per edge. The figure is based on the result presented in \cite{shmueli2015ride}.}
\label{fig:rides_dist}
\end{figure}

\begin{figure}[h!]
\centering
\includegraphics[bb=0 0 1500 1200, clip=true, scale =0.15]{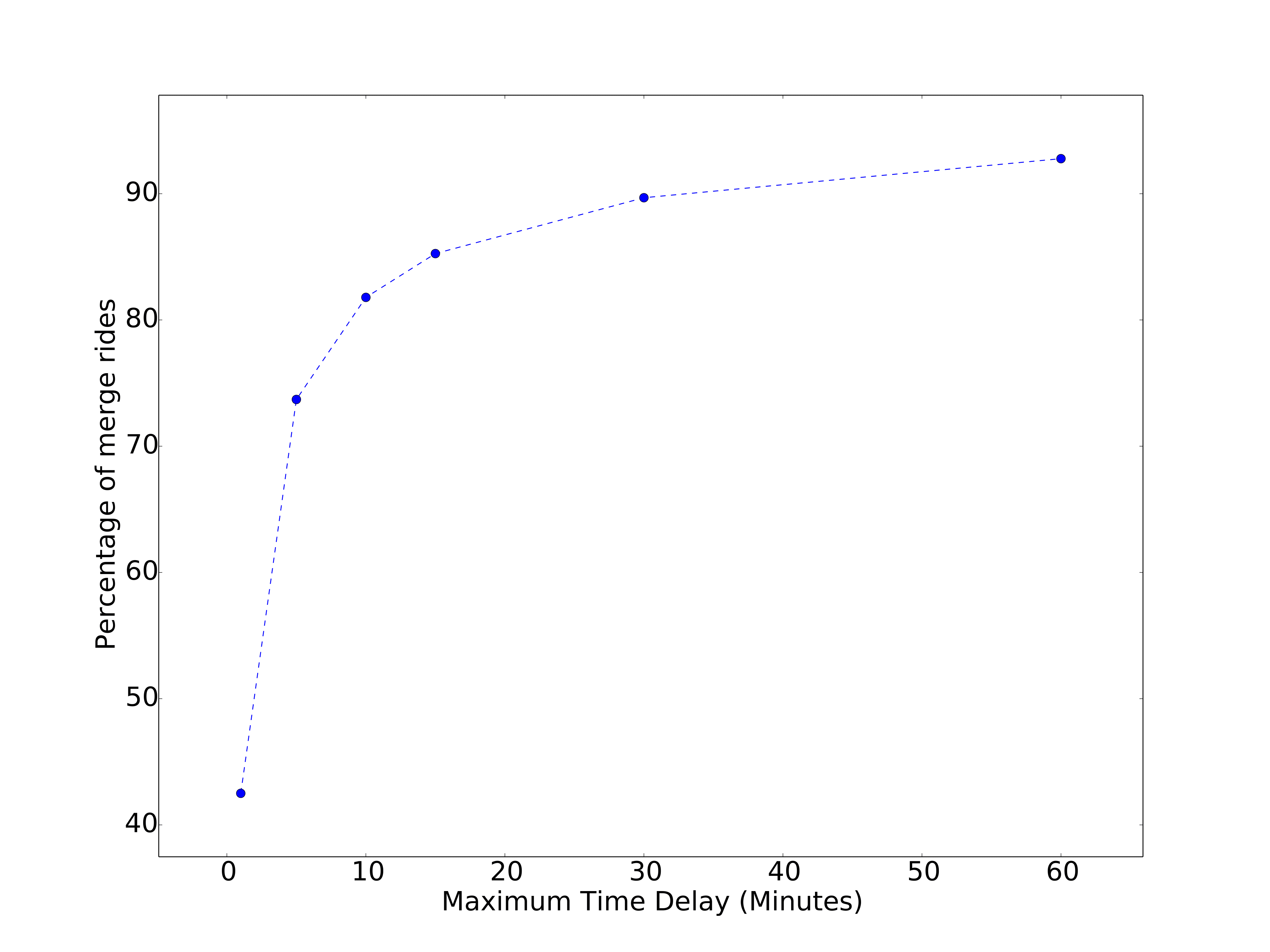}
\caption[Percentage of merged rides]{Percentage of merged rides (for the entire network). The figure is based on the result presented in \cite{shmueli2015ride}.}
\label{fig:monthly_benefit}
\end{figure}

Figure~\ref{fig:rides_dist} shows the probability density function (pdf) of the number of rides per edge.
As can be seen from the Figure, the distribution is heavy tailed and seems to follow a power-law.
In other words, most of the edges (i.e., pairs of origin-destination) induce a small number of rides while a small number of edges induce an extremely high number of rides.

Figure~\ref{fig:monthly_benefit} presents the percentage of shareable rides as a function of the maximum time delay parameter.
Results are encouraging: more than 70\% of the rides can be shared when passengers can accept a delay of up to 5 minutes.
As expected, the benefit of ride sharing increases when the passengers are willing to take a higher discomfort, and the percentage of shareable rides is more than 90\% when passengers can wait 30 minutes or more.

It should be noted that the simplified analysis illustrated in Figure~\ref{fig:monthly_benefit} assumes that two rides that took place at the same time can always be merged, regardless of the number of passengers in each ride.
Since the average number of passengers per ride is 1.7 and most of the rides involve a single passenger, the number of saved rides could have been even higher by merging more than 2 rides at a time.
On the other hand, in some cases, even the merging of two rides at a time might have resulted in overcrowding of the vehicle.

In order to assess the effect of these two potential phenomena over our analysis we can observe the distribution of the number of passengers per trip in the data. While doing so, we artificially segregate trip made using private taxi caps (that can board up to 4 passengers) and trips made with larger vehicles (capable of boarding from 5 to 48 passengers)~:

\begin{itemize}
  \item 49.22 percent of the trips have 1 passenger.
  \item 24.22 percent of the trips have 2 passengers.
  \item 15.72 percent of the trips have 3 passengers.
  \item 10.84 percent of the trips have 4 passengers.
\end{itemize}

We examine two approaches for the assessment of the actual theoretical ride-sharing utilization~:
\begin{description}
  \item[Greedy merging, assuming an even distribution of number of passengers] In this approach analyze the merging process in a two phases greedy approach. In the first phase, we assume that all the original trips that can be merged are indeed merged, and are done so under the assumption that the number of passengers is distributed approximately uniformly, with respect to the various geographic locations. Then, the resulting merged trips are merged again, if possible. This analysis approach should result in a lower bound for the actual ride-sharing utilization, as in real life our ride-matching algorithm would aspire for maximizing the number of merged rides, where possible.
  \item[Optimal merging] In this approach we assume that whenever two rides are merged, the number of passengers they have receives the value that would result in the most efficient merging scheme possible (confined to the overall distribution of the numbers of passengers for rides). This analysis approach should result in an upper bound for the actual ride-sharing utilization, as in real life there will be times where the only way to merge rides would be in a sub-optimal way.
\end{description}

Following is a detailed analysis of both approaches~:

\noindent\textbf{Greedy merging:}
The expected distribution of the merged trips for the first phase would be~:
\begin{itemize}
  \item In 24.23 percent of the pairs, we would merge a trip that has 1 passenger with a trip that has 1 passenger. This results in a merged trip of 2 passengers.
  \item In 23.84 percent of the pairs, we would merge a trip that has 1 passenger with a trip that has 2 passengers. This results in a merged trip of 3 passengers. These trips cannot be merged, assuming the greedy 2-steps approach.
  \item In 15.48 percent of the pairs, we would merge a trip that has 1 passenger with a trip that has 3 passengers. This results in a merged trip of 4 passengers, that cannot be further merged.
  \item In 5.87 percent of the pairs, we would merge a trip that has 2 passengers with a trip that has 2 passengers. This results in a merged trip of 4 passengers, that cannot be further merged.
  \item In 30.58 percent of the pairs, we would not be able to merge the trips, has these would be pairs that either (a) have one of the trips with 4 passengers, or (b) having a trip with 2 passengers and a trip with 3 passengers, or (c) having two trips having 3 passengers each.
\end{itemize}

The second phase will therefore be able to merge another $0.2423 \cdot 0.2423 \cdot 100 = 5.87$ percent of the original pairs, which reflects a $5.87 \cdot 2 = 11.74$ percent increase. Overall, this would sum up to $100 - 30.58 + 11.74 = 81.16$ percent of the naive potential utilization (namely, the utilization that is calculated under the assumption that all rides are merge-able, and that we do not merge more than two rides.

\noindent\textbf{Optimal merging:}
Assuming an optimal merging scheme we can calculate the merging of the relevant New York City data as follows~:

\begin{itemize}
  \item The 10.84 percent of the rides that have 4 passengers cannot be merged at all.
  \item The 24.22 percent of the rides that have 2 passengers would be merged among themselves.
  \item The 15.72 percent of the rides that have 3 passengers would be merged with a matching 15.72 percent of the rides that have 1 passenger.
  \item This would leave another (49.22 - 15.72 =) 33.5 percent of the rides, that have 1 passenger. These rides would be merged in a 4-to-1 ratio, virtually implying a $33.5 \cdot 1.5 = 50.25$ percent save.
\end{itemize}
Altogether, the actual optimal theoretical utilization would sum up to $24.22 + 15.72 + 15.72 + 50.25 = 105.91$ percent (namely, under the assumption of optimal merging the benefit from merging 4 rides of a single passenger more than compensates the loss due to rides with 4 passengers.

Therefore, the actual theoretical utilization for the New York City taxi dataset, denoted as $U$, would be bounded by~:
 \[0.8116 \cdot \alpha \leq U \leq 1.0591 \cdot \alpha\]
such that $\alpha$ is the potential utilization that is calculated throughout this work, using the method that was described above, ignoring the effect of multiple merges, as well as the effect of over-population of rides.

\section{Analyzing the Dynamic Ride-Sharing Network}
\label{sec:network}

In the previous Section we have described the taxi data that was used for this study, illustrated various mathematical properties of it, and discussed the way it can be analyzed for the purpose of assessing the potential ability of ride-sharing schemes to merge rides between similar locations (denoted as the \emph{ride-sharing potential utilization}).
In this Section we demonstrate the inability of static analytic approaches to efficiently model this utilization and suggest an alternative approach, that is based on the construction of multiple network-snapshots, derived using a sliding-window based aggregation of the taxi rides. We show that this technique can serve as a valuable methodology for both (a) assessing the potential ride-sharing utilization of the current supply and demand scheme (as appears in Section \ref{sec.dyn_analysis}), as well as (b) serve as a prediction method for estimating \emph{changes} in this utilization, in the near future, up to a few hours (as shown in Section \ref{sec.dyn_prediction}).

\subsection{The Need for Dynamic Ridesharing Optimization and Prediction}

Mainstream transportation analysis models (such as \cite{ben1985discrete,erlander1990gravity,wen2001generalized,transportation-Wardropuserequilibrium,mcnally2007four} and many more) approach the problem of transportation forecasting and analysis through the use of long-term data aggregation. Simply put, the dominating approach today sees the accurate approximation of the ``steady state'', or ``average state'', of the transportation system as the most efficient way to understand the behavior of the system, and to use this understanding in order to reach better decisions \cite{friesz1985transportation}.
Such decisions are often concerned with the locations, type or size of new infrastructures that should be built, large-scale budgets investment alternatives or long-term policy revisions \cite{meyer1984urban}.

When examining the rapidly expanding field of ridesharing this approach suffers an inherent limitation, as it not well adequate for the nature of decisions ridesharing operators and regulators are require to make. As ridesharing uses existing roads and metropolitan infrastructure, does not require setting fixed-place stations of fixed-paths, and often uses existing vehicles, it is mostly located ``outside'' the realm of these analysis methodologies. Furthermore, ridesharing introduces a new set of factors that traditional methods usually cannot easily cope with, such as dynamic changes in fares, which may significantly influence network properties such as global congestion \cite{xu2015traffic}.

Analyzing ridesharing using the existing models would be inefficient at best.
Taking the static approach using a long-term aggregation of the supply and demand would inevitably result in a model that would be optimized for the \emph{average states} of the rides network, ignoring its inherent volatility (that is caused due to daily and weekly patterns as well as irregular spikes created by events such as street-parties, sports events, etc.).

Interestingly, as shown in Section \ref{sec.dyn_analysis}, the dynamic rides network spends only an extremely \emph{small portion} of the time in those average network states. Furthermore, our analysis demonstrates that overlooking the dynamic nature of the traffic scheme disregards the vast majority of the network states, as manifested in the O-D matrix, as well as the possible ridesharing utilization of it.
Specifically, this phenomenon is demonstrated in Figure \ref{fig:dynamicStat1} that reveals that the system spends approximately 33\% of the time in states that have a potential utilization of either 50\% above the monthly average, or 50\% below it.

Ignoring this dynamic nature of the urban rides system through the use of a static analysis model (which is the mainstream approach of today) will be inherently limited in its efficiency. The key to unlocking the development of effective next generation ridesharing systems therefore lays in an analysis that is rooted in the understanding of its dynamic nature, and the way to use it in order to develop pro-active strategies that dynamically adapt their forecast using an ad-hoc analysis of the network's state.

A potential example for this approach can be found in \cite{lee2015dynamic}, containing a computational study aimed for identifying environments in which the use of ``dedicated drivers'' are most useful. As urban supply and demand environments are constantly (and significantly) changing (as demonstrated in our analysis of the New York taxi data), it is therefore likely that a strategy that detects the \emph{times} where the use of such drivers is most efficient and upon such detection -- launches these drivers to supply the demand\footnote{This can be done using a dynamic change in the commission drivers are required to pay, giving such drivers a temporary priority in certain roads, or forbidding them from granting service on a regular basis expect from when their service is required.} -- would achieve a superior performance compared to a static strategy that does not react to such changes.

Another example can be the work of \cite{fagnant2015dynamic} in which the size of a carsharing fleet is optimized in order to maximize the monetary operational savings. Again, such an approach reaches the global optimization assuming a static approach whereas the incorporation of the dynamic nature of the system could yield a significant. This could be done for example by allowing the fleet operators to dynamically use the services of a public service (such as Uber or Lyft), rented cars, or private drivers. Using such service when needed will allow to reduce the ongoing basic cost.

\subsection{Dynamic Network Analysis}
\label{sec.dyn_analysis}

As discussed in previous sections one of the main hurdles that prevent the wide adoption of ride-sharing might be the high volatility of its potential utilization, and the extreme unpredictability of it. In this section we propose to mitigate this problem by using a dynamic network that represent the evolving travel patterns in the city. That is, a multitude of rides-networks, representing data of fixed-length periods of time, each of which starting at different points in time of equal distances. Such ``sliding window'' approach is useful for tracking changes in various properties of this dynamic network, which we show are not only highly correlated with the potential ride-sharing utilization at the corresponding points in time, but can also \emph{predict} the utilization few hours ahead of time.

\begin{figure}[h!]
\centering
\includegraphics[bb=0 50 600 400, clip=true, scale =0.5]{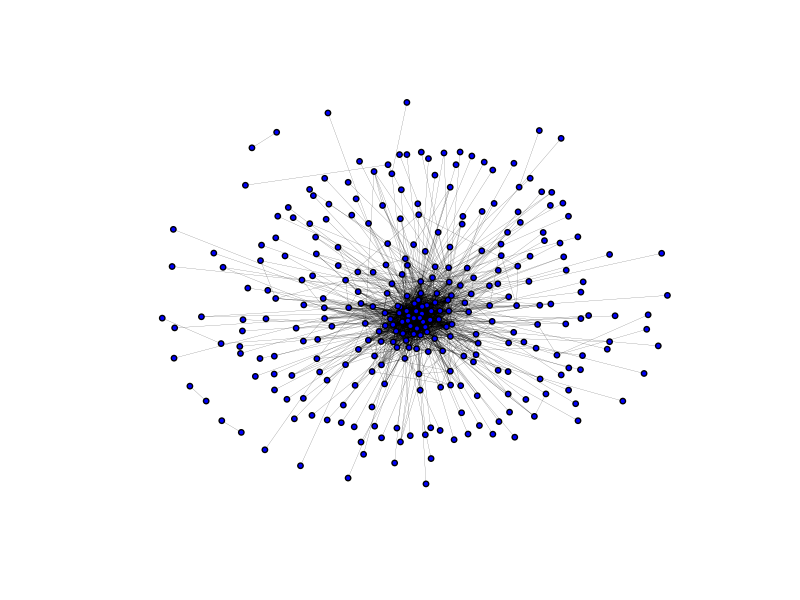}
\caption[An illustration of the rides sub-network $G_{T_{144},T_{145}}$]{An illustration of the rides sub-network $G_{T_{144},T_{145}}$, denoting the structure that is implied by the aggregation of the rides between the 144-th and the 145-th hour of the month.}
\label{fig:network}
\end{figure}

We divide the rides dataset into hourly aggregated snapshots, creating $31 \times 24=744$ sub-networks, each is denoted by $G_{T_{n},T_{n+1}}$, such that $T_{n}$ represents the $n$-th hour in the month.
An illustration of one such sub-network is shown in Figure \ref{fig:network}.
Intuitively we see that most of the nodes are highly connected, but a considerable number of nodes are connected to only one other node in the network.

\begin{figure}[h!]
\centering
\includegraphics[bb=0 0 1500 1200, clip=true, scale =0.15]{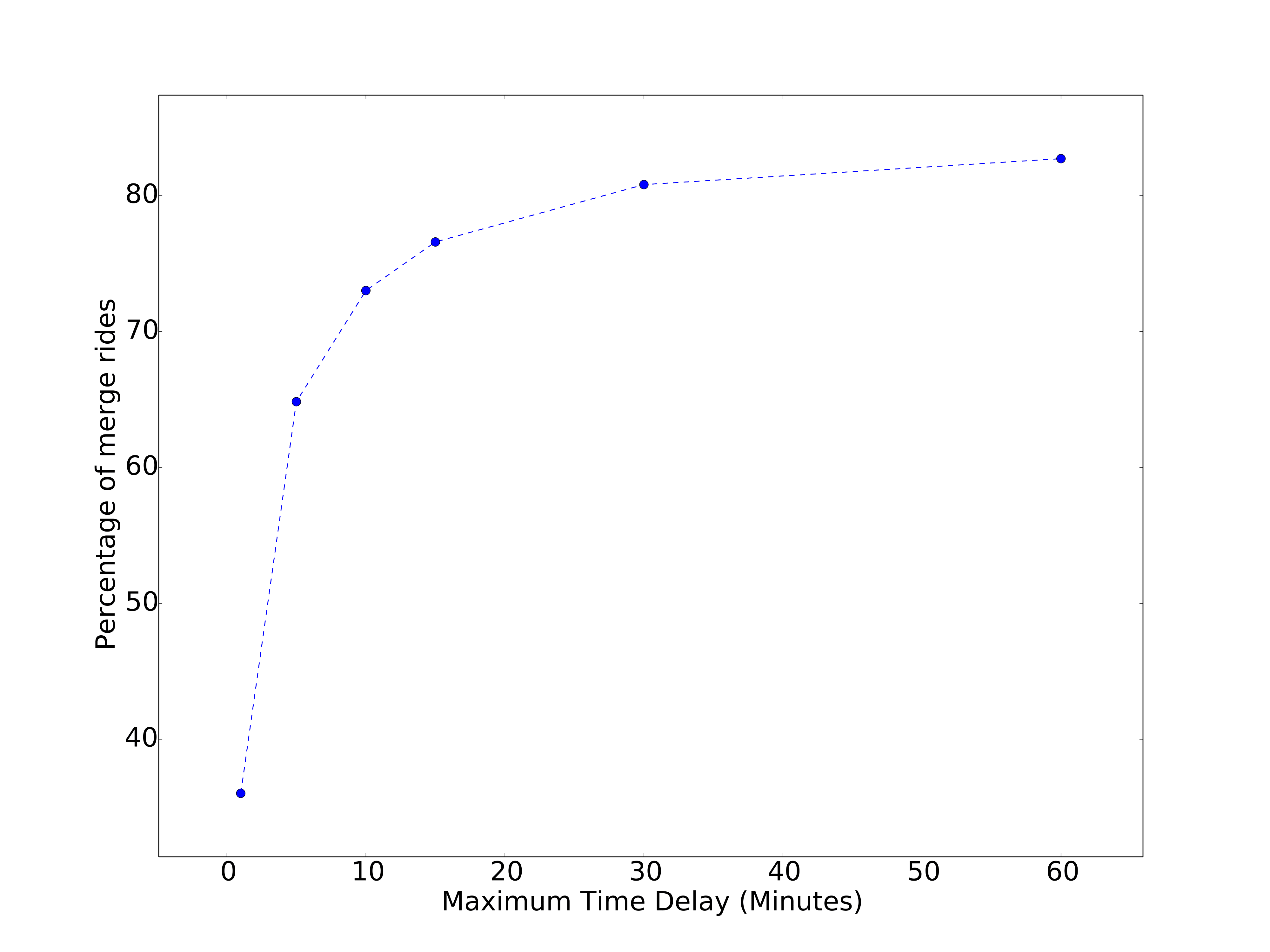}
\caption[Potential of ride-sharing utilization, measured as the average percentage of potentially merged rides]{Potential of ride-sharing utilization, measured as the percentage of potentially merged rides (averaged over all sub-networks), as a function of maximal delay agreeable by the passengers. The figure is based on the result presented in \cite{ICCSS15Altshuler}.}
\label{fig:hourly_benefit}
\end{figure}

Similarly to Figure~\ref{fig:monthly_benefit} in which the potential benefit of ride-sharing over the entire data was shown,
we have performed the same calculation for every hourly network separately. Figure~\ref{fig:hourly_benefit} presents the average potential ride-sharing utilization taken on all hourly networks, as a function of the maximal delay allowed (notice that this is in fact a lower bound, since we artificially prevent passengers from being merged with rides ``outside'' their hourly network). It can be seen that this produces a lower utilization than the previous calculation using the overall aggregation (approximately 10\% decrease), caused by the fact that each pair of nodes has a lower probability of being connected.

\remove{This result is expected as each sub-network contain less rides for each edge and therefore there are less options for merging rides.
Nevertheless, the curves have a similar trend in the two scenarios.
This suggests that the effect of the quality of service is similar even on different networks.
}

We now extract a set of six common network properties for each traffic-network $G_{T_{i},T_{j}}$, to be used as the features values representing each network. These features encapsulate various topological aspects of the network and enable us to project each hourly-collection of traffic data (containing a large and apriorically unknown number of rides) into a single coordinate in a 6-dimensional feature-space.

\subsubsection{(1) Number of Nodes} The number of nodes in the network $G$, denoted as $|V|$, representing the number of unique pick-up and drop-off locations of rides made during this time window. Note that although all the networks refer to the same dataset, and the same geographic environment, different networks may have different values of $|V|$, since at different time-segments different locations may be ``active''.

\subsubsection{(2) Number of Edges} The number of edges in the network $G$, denoted as $|E|$, representing the number of unique pick-up to drop-off pairs of rides made during this time window. This is also the number of non-zero elements of the temporal O-D matrix that is derived from this network.

\subsubsection{(3) Network Density} The average degree of the network's nodes, defined as $\frac{|E|}{|V|}$.
This property represents the average number of unique drop-off locations per pick-up location (and vice versa) and is associated with the predictability of rides made during this time window, and is also related to the system's entropy.

\subsubsection{(4) Average Betweenness Centrality} Each node $v$ in the network $G$ has a calculate-able betweenness centrality score \cite{BCDef1}, representing the portion of ``shortest paths'' between all the node-pairs in the network, that pass through $v$. Formally, for a network node $v \in V$ this is defined as~:
 \[\sum_{s \neq v \neq t} \frac{\sigma_{s,t}(v)}{\sigma_{s,t}}\]
where $\sigma_{s,t}$ is the total number of shortest paths from node $s$ to node $t$ and $\sigma_{s,t}(v)$ is the number of those paths that pass through $v$.

Averaging these values yields an estimation of the network's efficiency, with respect to the number of nodes whose adequate availability is required in order to preserve the network's ability to maintain efficient flow without increasing the length or durations of trips between arbitrary points \cite{puzis2012augmented,altshuler2011augmented}.

\subsubsection{(5) Average Closeness Centrality} The closeness centrality of a node \cite{stephenson1989rethinking} is a measure of centrality in a network, calculated as the sum of the length of the shortest paths between the node and all other nodes in the graph. Thus the more central a node is, the closer it is to all other nodes. For a node $v \in V$, the measure is defined as~:
\[\frac{1}{\sum_{x} d(v,x)}\]

Averaging the closeness centrality over all the network's nodes yields an estimation of the compactness of the network, that is -- how short it is to travel between an arbitrary pair of network nodes.

\subsubsection{(6) Average Eigenvalue Centrality} Eigenvalue centrality \cite{bonacich1972factoring} (also called eigencentrality or eigenvector centrality) is a measure of the influence of a node in a network. It assigns relative scores to all nodes in the network based on the concept that connections to high-scoring nodes contribute more to the score of the node in question than equal connections to low-scoring nodes.

For a given graph $G$ with an adjacency matrix $A$ the centrality score of a node $v \in V$, denoted as $x(v)$, is defined as
\[\frac {1}{\lambda} \sum_{u \in M(v)}x_{u}\]
where $M(v)$ is a set of the neighbors of $v$ and $\lambda$ is the graph's largest positive real eigenvalue.
This can be accurately estimated by taking the $v^{th}$ component in the eigenvector that corresponds to the largest positive real eigenvalue.

The use of eigenvalues to analyze propagation phenomena over networks can be see for example in \cite{prakash2010got}, where its usability for predicting the epidemic potential of viruses is demonstrated.

\begin{figure}[h!]
\centering
\includegraphics[bb=-100 0 2000 480, clip=false, scale=0.4]{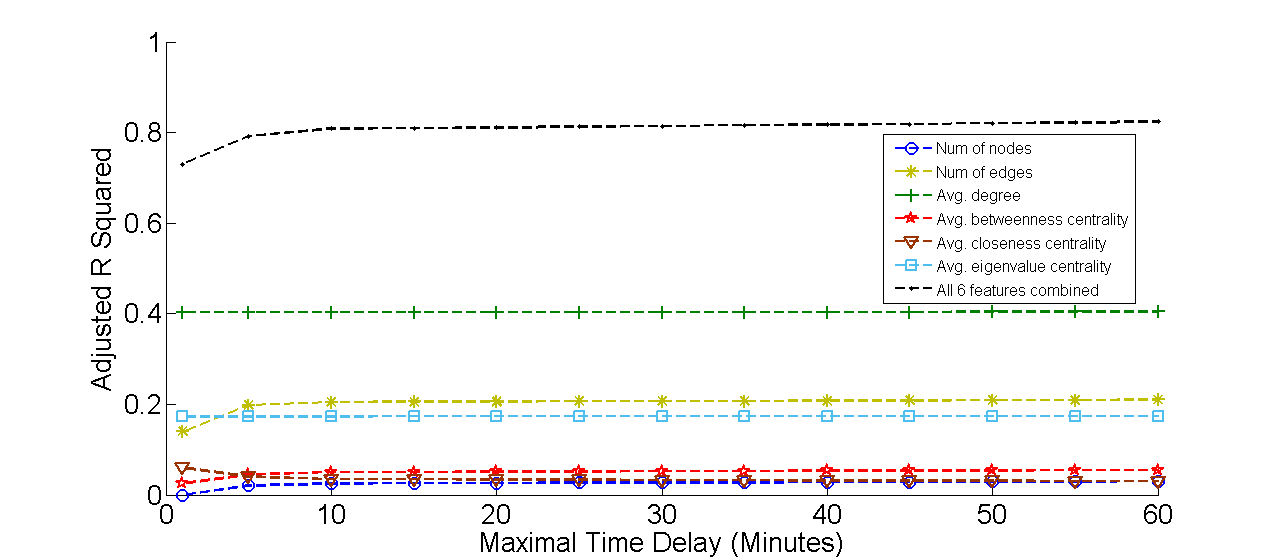}
\caption[Adjusted $R^2$ of the correlation between features of the hourly rides network and the potential ride-sharing utilization]{Adjusted $R^2$ of the correlation between seven features of the hourly rides network and the potential ride-sharing utilization for this network. Most features have low quality of fit, but the combined mixture of all seven results in a remarkably high correlation ($R^2 = 0.82$).
Features are (1) the number of nodes in the network, (2) the number of edges, (3) the averaged degree, (4) the averaged betweenness centrality, (5) the averaged closeness centrality, and (6) the averaged eigenvector centrality.}
\label{fig:hourly_r2}
\end{figure}

We use a linear regression to fit these features for the calculated potential utilization, as well as a multiple linear regression to fit the potential utilization for the entire set of network properties.
As can be seen in Figure~\ref{fig:hourly_r2} these features show a high correlation with the potential utilization for this hourly network (the figure reports the adjusted $R$ squared to account for the different number of predictors).

\subsection{Ride-Sharing Potential Prediction}
\label{sec.dyn_prediction}

In the previous section we have shown that the monthly rides can be partitioned into hourly aggregative snapshots, each of different characteristics (and specifically, network oriented ones), and different ride-sharing potentials. In addition, we have demonstrated the correlation between these network properties and the ride-sharing potentials of the rides the corresponding networks are implied from (as appears in Figure~\ref{fig:hourly_r2}).
In this section, we discuss whether this correlation can also be used for predictive purposes. Specifically, can we deduce from the current values of various network properties how the \emph{change} in the ride-sharing potential compared to its current value.

\begin{figure}[htbp]
\centering
\includegraphics[bb=-100 0 1200 600, clip=true, scale=0.4]{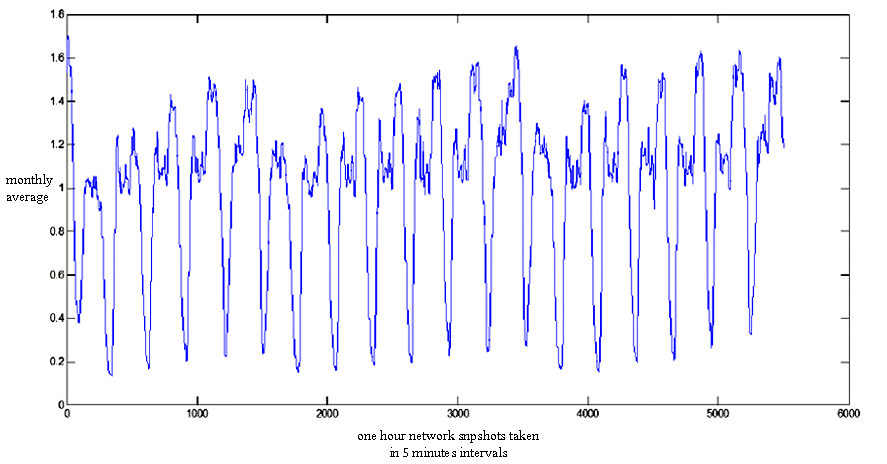}
\caption[Dynamics of the mean degree of the rides network nodes]{Dynamics of the mean degree of the rides network nodes. X-axis denotes the time, given in 5 minutes granularity. Y-axis denotes the change of the mean degree of the network compared to its monthly average.}
\label{fig:dynamicStat2}
\end{figure}

In order to do so,  we first analyze the the evolution of various network properties of the hourly aggregative rides network $G_{T_{n}, T_{n+1}}$ over time. Figure \ref{fig:dynamicStat2} illustrates the evolution of the mean nodes degree of the rides network as a function of time (that is, the average over all of the network's nodes' degrees, for all the dynamic hourly networks). For the sake of clarity, we have increased the time granularity used in the analysis, so that the hourly networks are now generated with 5 minutes intervals, thus significantly overlapping, and subsequently generating a smoother and easier to read graph. The change from the monthly average of the mean degree as a function of time is portrayed, clearly showing a dominant daily pattern. However, on top of this pattern we can see significant hourly fluctuations, tens of percent in magnitude. This reveals the existence of strong volatility in the rides dynamics alongside the predicted daily and weekly dynamics.

\begin{figure}[htbp]
\centering
\includegraphics[bb=-100 0 1200 600, clip=true, scale=0.3]{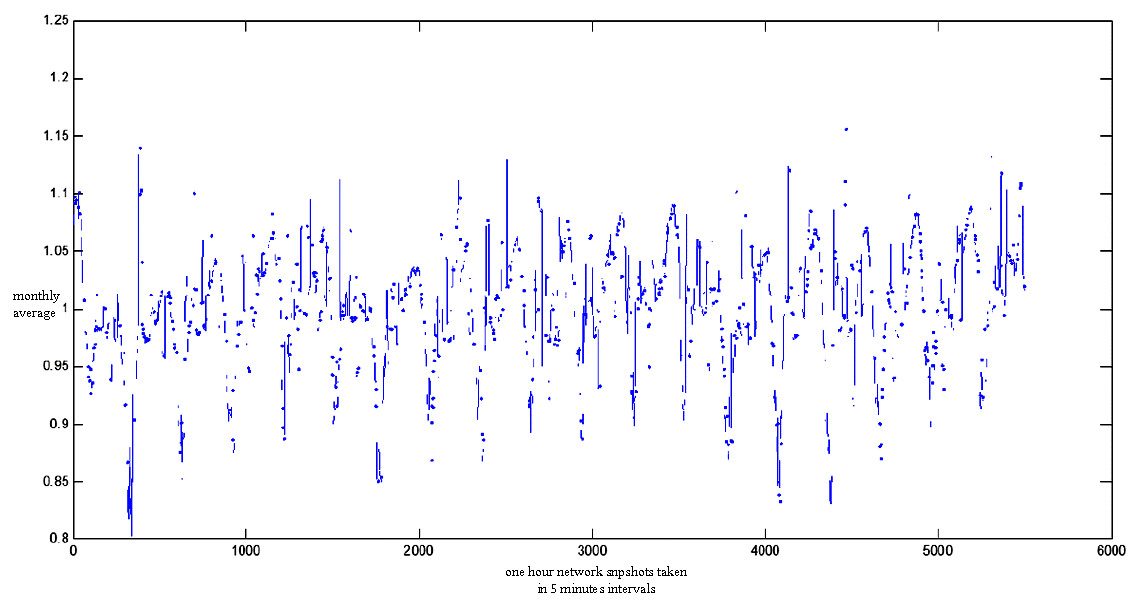}
\caption[Dynamics of the largest eigenvalue of the rides-dynamic network over time]{Dynamics of the largest eigenvalue of the rides-dynamic network over time. X-axis denotes the time, given in 5 minutes granularity. Y-axis denotes the change of the largest eigenvalue of the network compared to its monthly average.}
\label{fig:dynamicStat3}
\end{figure}

A similar dynamics is observed when analyzing the evolution of the largest eigenvalue of the rides-network' adjacency matrix over time.
The use of eigenvalues to analyze propagation phenomena over networks can be see for example in \cite{prakash2010got,Epidemic-Chakrabarti}, where its usability for predicting the epidemic potential of viruses (both human and computer-based) is demonstrated. Additional mathematical analysis on the role of eigenvalues in the analysis of network structures can be found in \cite{boccaletti2006complex}.
This property, known to encapsulate various behavioral characteristics of the people whose mobility patterns the network is depicting, displays a clear (and easy to predict and understand) daily pattern, on top of which significant and erratic spikes are added, as can be seen in Figure \ref{fig:dynamicStat3}. These spikes seem to appear sporadically, lacking any clear patterns or internal regularity, implying again the need for understanding the dynamic aspects of the network.

\begin{figure}[htbp]
\centering
\includegraphics[bb=-100 0 1200 600, clip=true, scale=0.3]{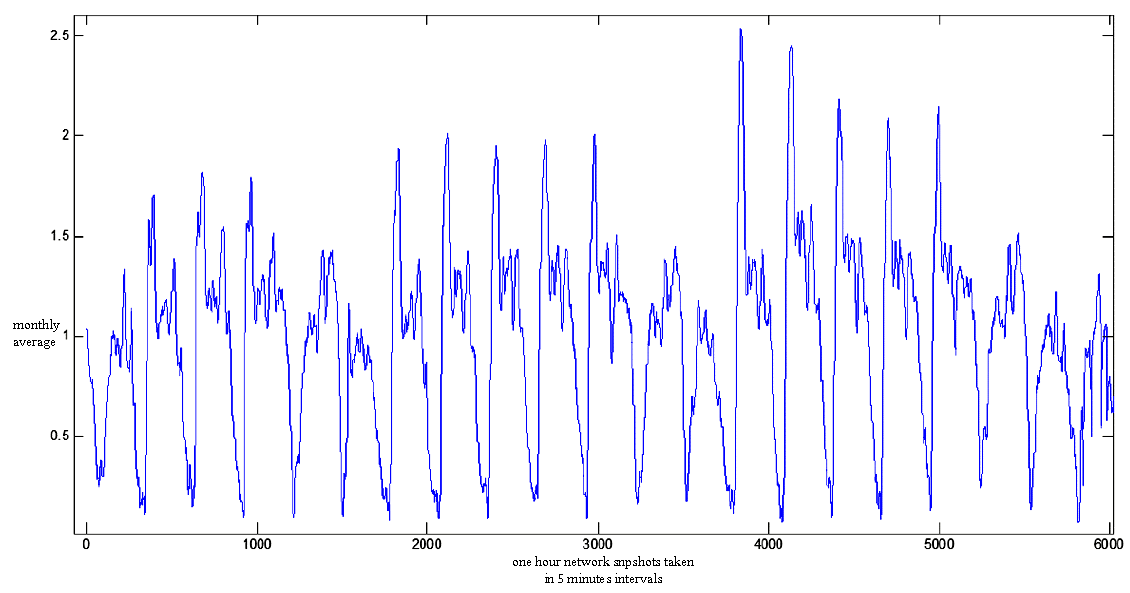}
\caption[Dynamics of the potential ride-sharing utilization over time]{Dynamics of the potential ride-sharing utilization over time. X-axis denotes the time, given in 5 minutes granularity. Y-axis denotes the change of the potential utilization compared to its monthly average.}
\label{fig:dynamicStat1}
\end{figure}

Now, let us perform a similar analysis over the potential ride-sharing utilization, looking at its evolution over time. The results of this analysis, presented in Figure \ref{fig:dynamicStat1}, clearly demonstrate a similar dynamics to the couple of network properties mentioned earlier. Specifically, it can be seen that alongside the dominating daily pattern (and weaker, but still easy to see, weekly one), there are clear changes in the potential utilization. These changes take various shapes and forms, from sudden decrease in the daily peak (as can be seen around $X=1400$), to changes in the intra-weekly peaks (the first week analyzed showing a `U-shaped' form among its days, the second week showing an equal-peaks dynamics, and the third week showing an extremely high Monday and Tuesday, and weaker Wednesday, Thursday and Friday), and others.
Surprisingly, the magnitude of these changes may even exceed the dominating daily pattern. For example, the change between the first Tuesday (around $X=200$) and the third Tuesday ($X=4100$) is 90\% compared to the monthly average, whereas the average change in potential utilization between workdays and weekends is only 70\%.

At this point we ask the following question: ``can we find a statistical correlation between \emph{current} values of the rides network properties and \emph{future} values of the potential ride-sharing utilization?''. This question is of interest, as such a correlation would allow us to predict future changes in the potential utilization, providing valuable tools for both ride-sharing users, operators, and regulators.

\begin{figure}[htbp]
\centering
\includegraphics[bb=-100 0 1200 600, clip=true, scale=0.3]{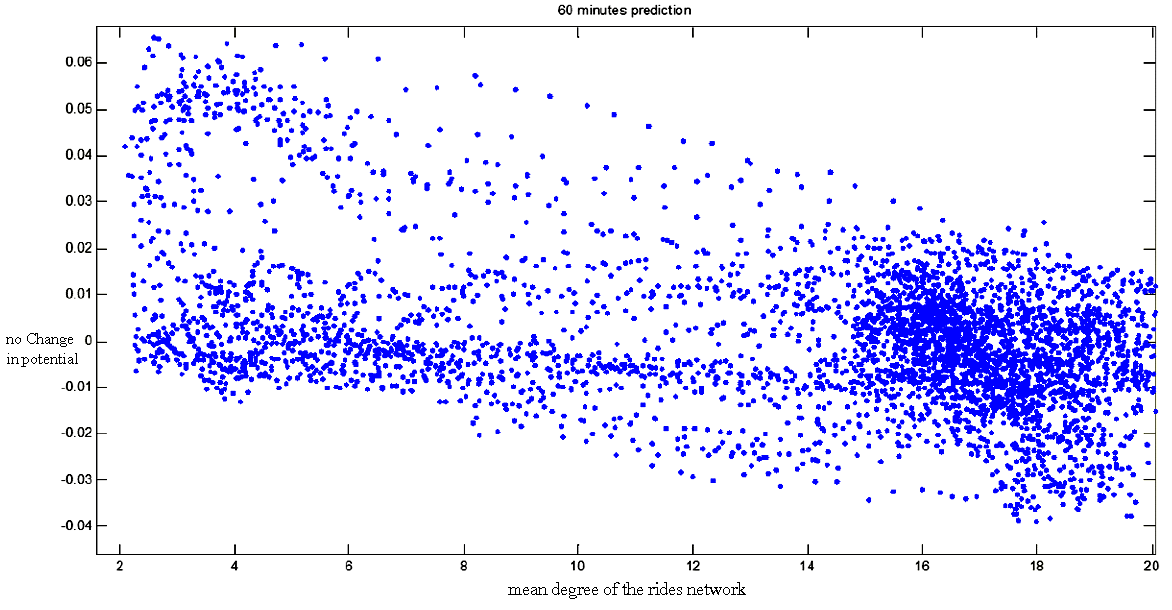}
\caption{Change in potential ride-sharing utilization (Y-axis), 1 hour prediction, as a function of the mean degree of the rides network (X-axis).}
\label{fig:dynamicStat5}
\end{figure}

We first address this question by comparing network properties values at time $T$ with potential utilization of at time $T+1$ (1 hour prediction). Figure \ref{fig:dynamicStat5} presents an example of such a comparison, in the form of a scatter plot showing for each point in time $T$ a dot whose X-axis is the mean nodes degree of the network $G_{T,T+1}$ and whose Y-axis is the change in the potential utilization of the rides between $T+1$ and $T+2$ compared to the rides between $T$ and $T+1$. That is, the change in the momentary ride-sharing utilization between ``now'' (time $T$) and ``in an hour'' (time $T+1$).
It is easy to see that this representation reveals a clear and strong negative correlation between the two.

\begin{figure}[htbp]
\centering
\includegraphics[bb=-100 0 1200 600, clip=true, scale=0.4]{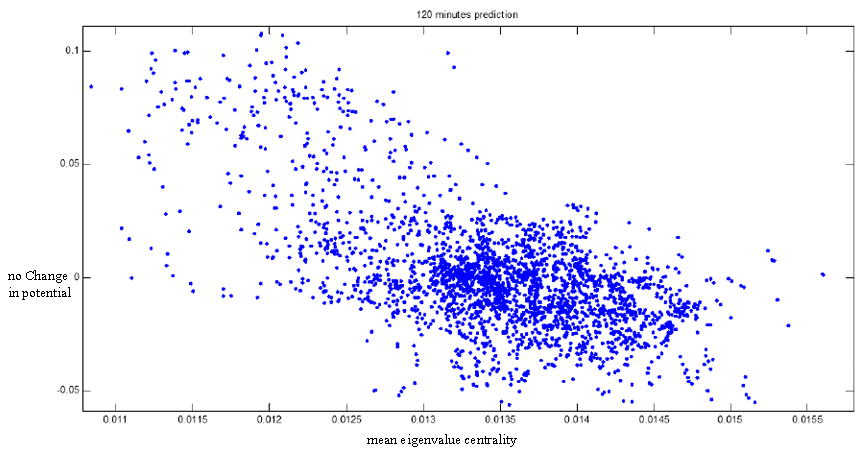}
\caption{Change in potential ride-sharing utilization (Y-axis), 2 hours prediction, as a function of the largest eigenvalue of the rides network (X-axis).}
\label{fig:dynamicStat4}
\end{figure}

Trying to increase our lookahead and predict the change in the dynamic ride-sharing utilization from a 2 hours time-frame,  Figure \ref{fig:dynamicStat4} illustrates the correlation between the value of the largest eigenvalue of the rides network at time $T$ and the change in the potential utilization between time $T$ (aggregated to $T+1$) and $T+2$ (aggregated to $T+3$).
Again, a clear strong negative correlation is easily visible. For example, in times where the value of the largest eigenvalue of the rides network is smaller than 0.012 the potential ride-sharing utilization was statistically guaranteed (during the month of the observation) to significantly increase in the coming 2 hours. Similarly, largest eigenvalue of 0.014 would indicate a significant decrease in the ride-sharing potential within the next 2 hours.

\begin{figure}[htbp]
\centering
\includegraphics[bb=100 0 1800 600, clip=true, scale=0.4]{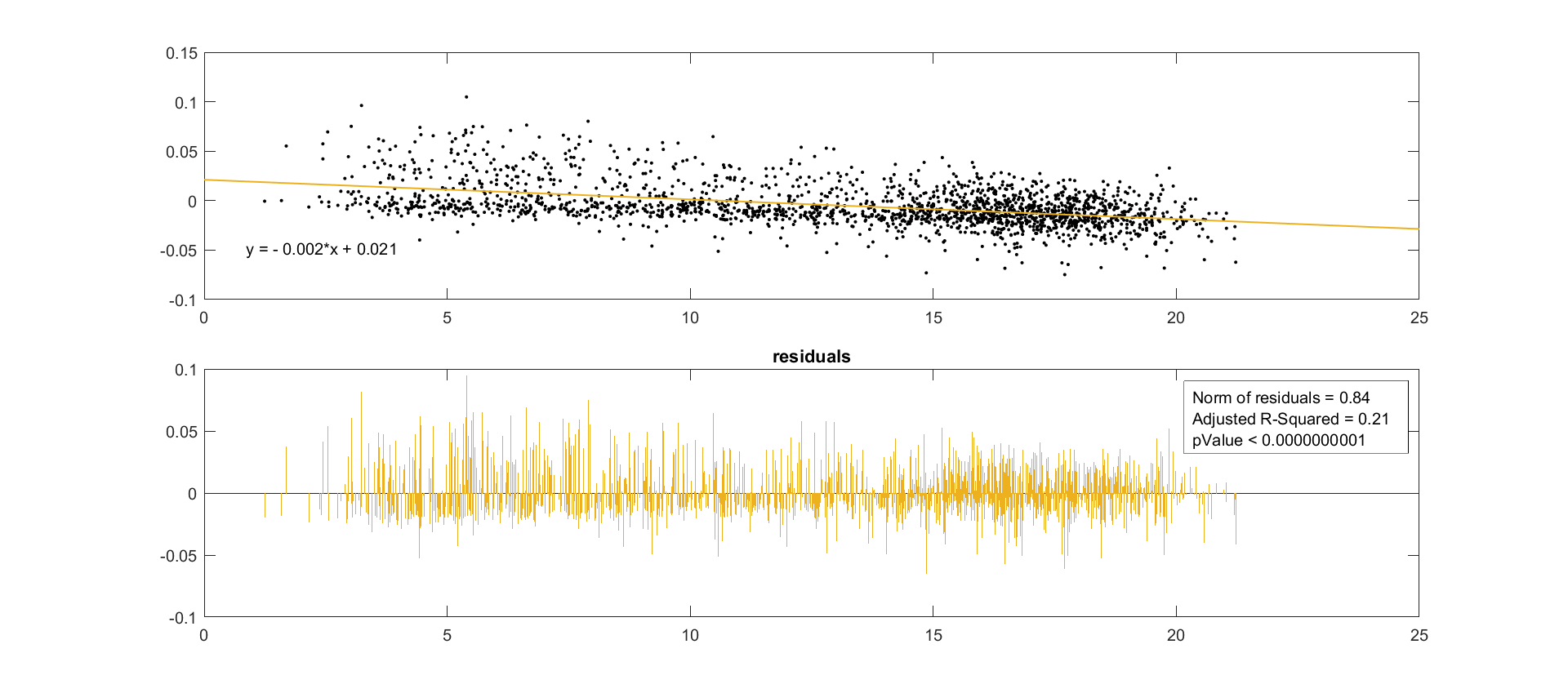}
\caption{Change in potential ride-sharing utilization (Y-axis), 1 hour prediction, as a function of the mean degree of the rides network (X-axis), created for the last week of the data.}
\label{fig:dynamicStat5_validation}
\end{figure}

\begin{figure}[htbp]
\centering
\includegraphics[bb=100 0 1800 600, clip=true, scale=0.4]{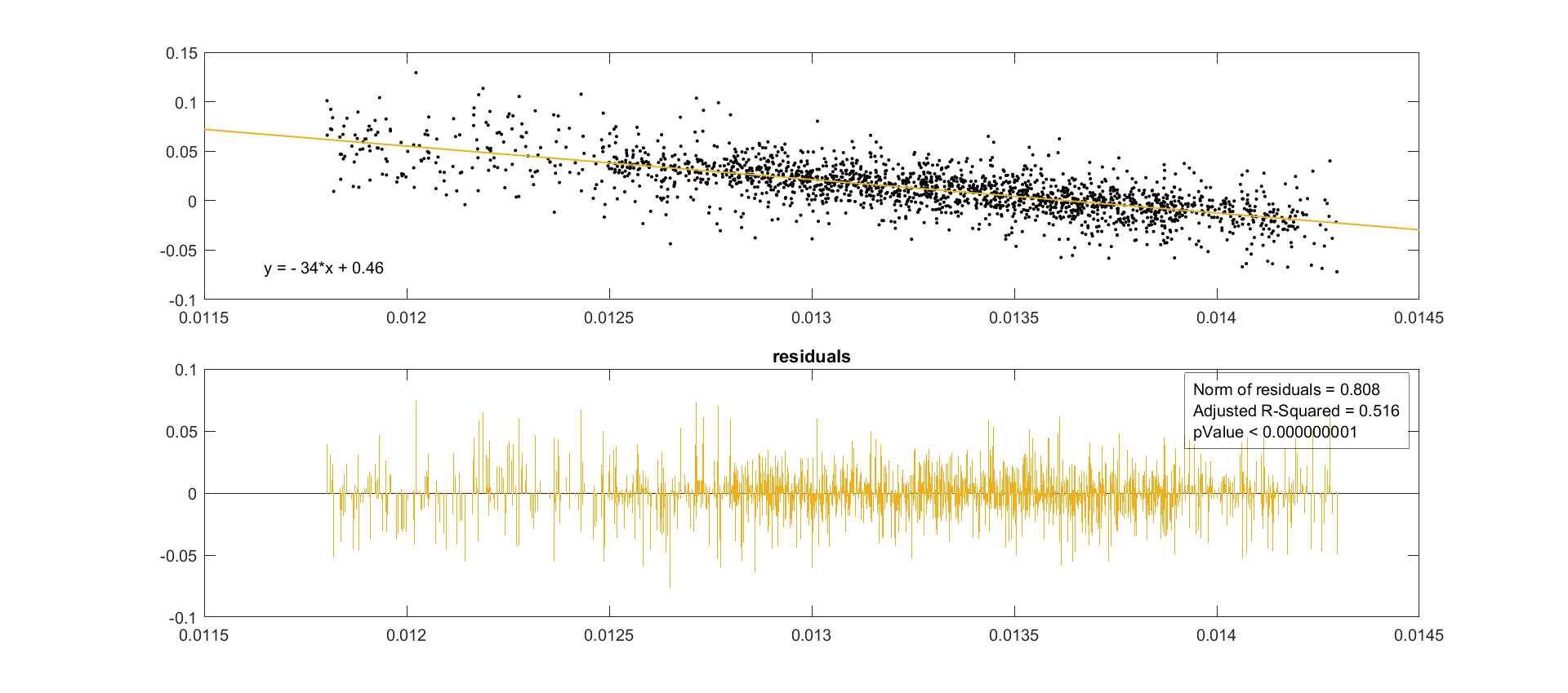}
\caption{Change in potential ride-sharing utilization (Y-axis), 2 hours prediction, as a function of the largest eigenvalue of the rides network (X-axis), created for the last week of the data.}
\label{fig:dynamicStat4_validation}
\end{figure}

Figures \ref{fig:dynamicStat5} and \ref{fig:dynamicStat4} are based on the analysis of the first 3 weeks of January 2013. These observations were then validated using the last 1 weeks of January, as can be seen in Figures~\ref{fig:dynamicStat5_validation} and \ref{fig:dynamicStat4_validation}.

Once demonstrating the predictive power of the dynamic network's properties with respect to the network's future ride-sharing potential we can now construct a  multiple linear regression model that would fit all of these 6 properties. We have created 18 models, for 2 values of distance tolerance (400 meters and 800 meters, denoting the pick-up and drop-off distances that still allow rides to be merged), 3 values of time tolerance (30 seconds, 2 minutes and 5 minutes, denoting the time passengers would be willing to wait in order to merge their rides) and 3 values of prediction horizon (no prediction, 1 hour prediction and 2 hours prediction). The results include a scatter plot of the data, effects of the various properties, Anova analysis and other statistical analysis appear in Figures \ref{fig:multilinear10} through \ref{fig:multilinear18}.

The effectiveness of the prediction as a function of the prediction horizon (i.e. the distance between the point in time where the prediction is calculated and the point in time this prediction refers to) is illustrated in Figures \ref{fig:multilinear_squarer1} through \ref{fig:multilinear_squarer6}, showing the $R^{2}$ of the model (both ordinal and adjusted) as a function of the time horizon (between 0 and 12 hours), for several values of distance tolerance and time tolerance. It can clearly be seen that in general (and as expected) the accuracy of the model decreases with the increase in the prediction horizon used (that is, when the model tries to predict the behavior of the system further into the future).

\section{Summary and Future Work}
\label{sec:summary_future}

As the popularity of ride-sharing systems grow, its users-base gradually transform from early adopters to mainstream consumers. Whereas the first are characterized by a keen affection for innovative solution that are powered by cutting edge technologies and aim to disrupt the governing paradigm in the field, the latter are often interested mainly in the advantages these services can offer them with as smallest change in their habits as possible. With respect to ride sharing these new users are willing to sustain far less wait-time and are extremely more susceptible to inconvenience than their preceding tech-savvy innovation-hungry early users. The key to a scalable mature ride sharing infrastructure is there found in the level of service such systems will provide, mainly measured by the availability of vehicles when they are needed. Alas, the availability maximization is immediately linked to a reduction in the financial savings that the service can offer.
In other words, a further expansion of ride-sharing is being constrained among others by the ability to offer high utilization, defined as the ability to ``merge'' similar rides in a way that would not require the passengers to sustain more than a minimal delay in their trips.

This optimization problem was extensively discussed in previous literature (comprehensive literature review can be found in Section \ref{sec:related}). However, the conventional approach to this problem assumed a static environment which needs to be optimized. By finding the optimal number of cars, or optimal pricing policy, the efficiency (or potential) of the system was assumed to be calculable in a robust way -- a key component in the decision of operators where to deploy new systems, in the design of relevant urban legislations by municipal policy makers, and of course in the likelihood of passengers to use these services.

In this work we discussed the dynamic nature of ride sharing systems. Specifically, we were interested whether ride-sharing utilization is stable over time (which coincides with the implicit assumption of most previous works in this field) or does it undergo significant and often rapid changes (which would imply the inherent inefficiency of schemes assuming a static nature). We modeled the ride sharing utilization using the known New York Taxi dataset and clearly shown that it is highly dynamic, and that any system that would be designed for the ``average'' utilization would be highly inefficient.

We then shown that assuming a dynamic approach the taxi data can be modeled as a sequence of data-snapshots, resulting in a dynamic traffic-network model. Several recent works jave shown that network features can effectively be used to predict a variety of events and properties, e.g., emergency situations, individuals' personality and spending behaviors~\cite{altshuler2013social,de2013predicting,singh2013predicting}. We used a similar technique in order to project the taxi data as into a feature space comprised of topological features of the dynamic network implied by this traffic. This (dynamic) feature space is then used to model the dynamics of ride-sharing utilization over time.

Using this approach we were able to demonstrate a clear correlation between the utilization of the ride sharing system over time and several topological features of the network it creates. In addition, we demonstrated that the potential benefit of ride sharing expressed as the percentage of rides that can be shared with a limited discomfort for riders can also be predicted a few hours in advance.
Such prediction can be used as a tool for an accurate short-term forecasting of the ride-sharing potential in cities and metropolitan areas.

Researchers in \cite{cici2014assessing,ma2013t,santi2014quantifying} and others have focused on addressing the computational challenges of trip-matching (an NP-hard optimization problem) in real-time and developed heuristics to quantify potential ride-sharing demand. These algorithms re-route trips in order to match
them with similar, overlapping trips, explicitly capturing demand for ridesharing relative to passenger's willingness to
experience prolonged travel time. However, finding an optimal solution to this problem is not computationally plausible (even under extreme limitations of the problem's space \cite{garey1976some}), and even the calculation of approximation heuristics would be computationally intense when done ad-hoc. Therefore, the ability to use current traffic dynamics in order to predict properties of an efficient near-future ride-sharing scheme -- such as the method we propose in this work -- can be used to make this process significantly more efficient \cite{kuipers2005conditions,naveh2007workforce}.

Future work should focus on the analysis of the correlation we find in this paper, trying to detect traces of possible causalities. Are network properties merely correlated with ride-sharing utilization, or do they possess an active influence over it? Evidences of the latter would enable us to offer urban designers and policy makers an innovative tool for encouraging and facilitating the adoption of ride-sharing systems.
Alternatively, incentives and fees could be better moderated, used as ``remedies'' in the case of a change in the travel patterns, in order to balance it and maintain a sustainable ride-sharing paradigm.
Another approach could be the pipelining of the dynamic ride-sharing utilization forecast as the input of models intended to predict the benefits of ride-sharing on the overall traffic \cite{bahat2015incorporating}.

Recent works have demonstrated the benefit of tracking the network's dynamics in order to improve collaborative decision making \cite{de2014strength,lazer2007network}. A possible continuation of the current work can analyze ride-sharing optimization as a case of decentralized decision making process, using the technique that is presented here.

As the prediction of future ride-sharing potential is ultimately needed for optimization purposes (of the overall travel time, congestion or any other utilization metric) of a dynamic coverage problem, comparing the performance of any proposed method to the theoretical results that are available for various types of such decentralized collaborative coverage challenges (see \cite{UAV-ROBOTICA,Altshuler-TCS2011,Svennebring1,Koenig2,new-IJRR,regev2012cooperative} and specifically \cite{altshuler2016optimal}) can also be of value.

Finally, as our suggested approach is agnostic to the actual route taken by the drivers it would be interesting to see whether the introduction of ride-sharing affects additional factors such as detours (that for a merged ride may become cost-effective), usage of toll-routes, etc.

\bibliographystyle{elsarticle-num}

\bibliography{Dynamic_ride_sharing_paper.bbl}

\clearpage
\appendix
\label{appendix}
\center
\huge
\textbf{Appendix A:}
\textbf{Prediction Results}


\begin{figure}[htbp]
\centering
\includegraphics[scale=0.4]{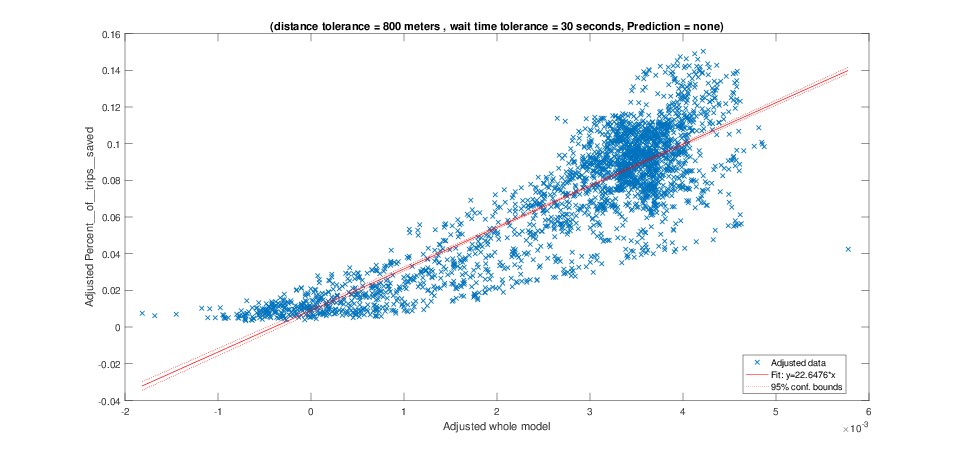}
\includegraphics[scale=0.4]{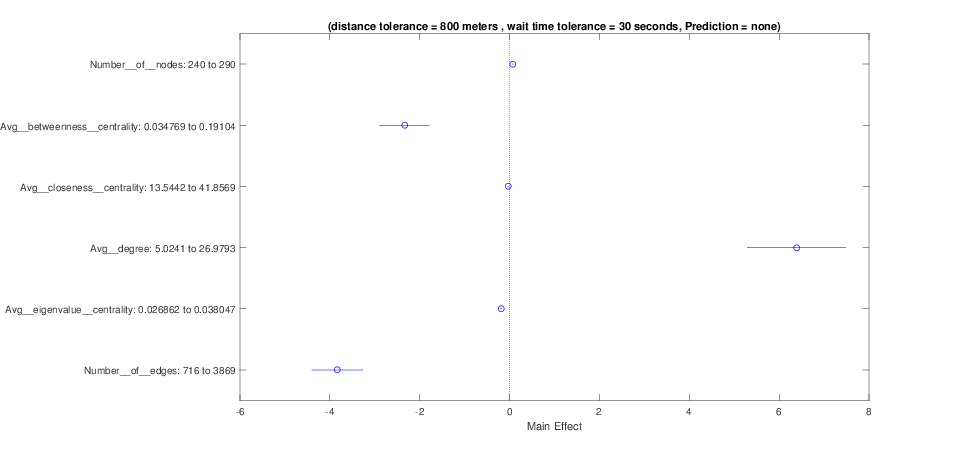}
\includegraphics[bb=0 0 850 700, clip=true, scale=0.35]{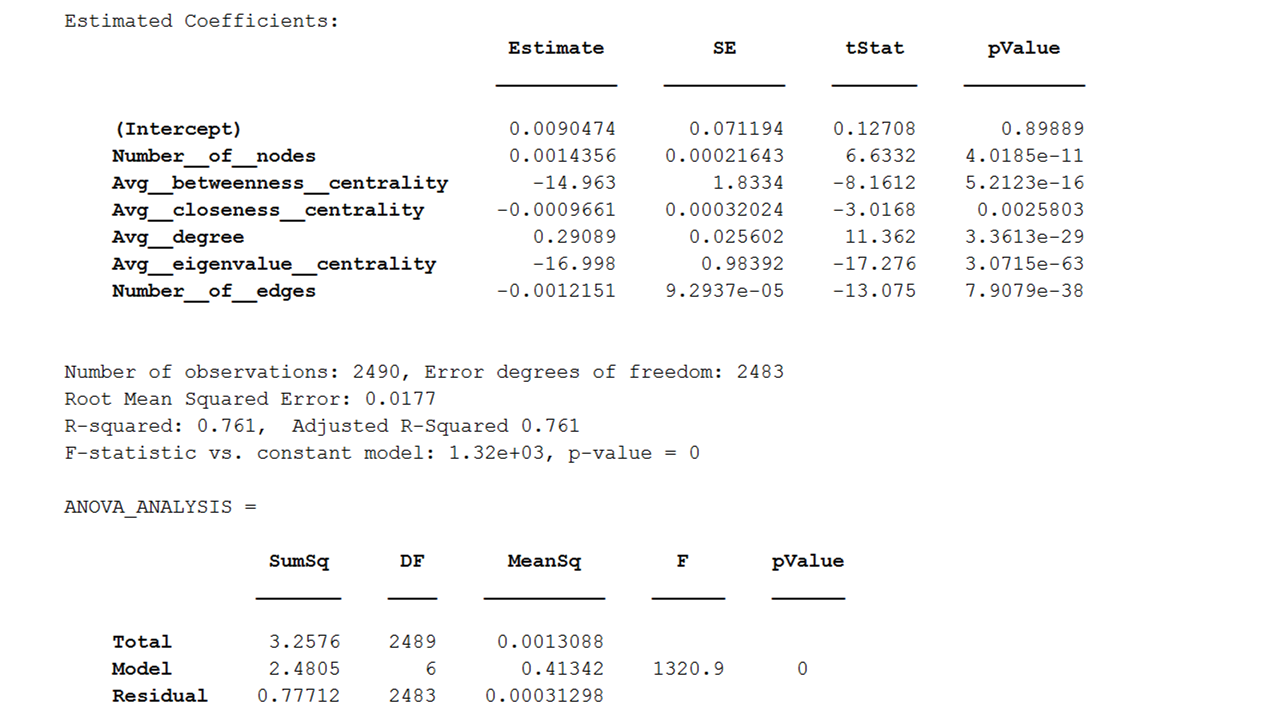}
\caption{Multilinear regression model, predicting the ride-sharing utilization using the dynamic network's properties 10.}
\label{fig:multilinear10}
\end{figure}

\begin{figure}[htbp]
\centering
\includegraphics[scale=0.4]{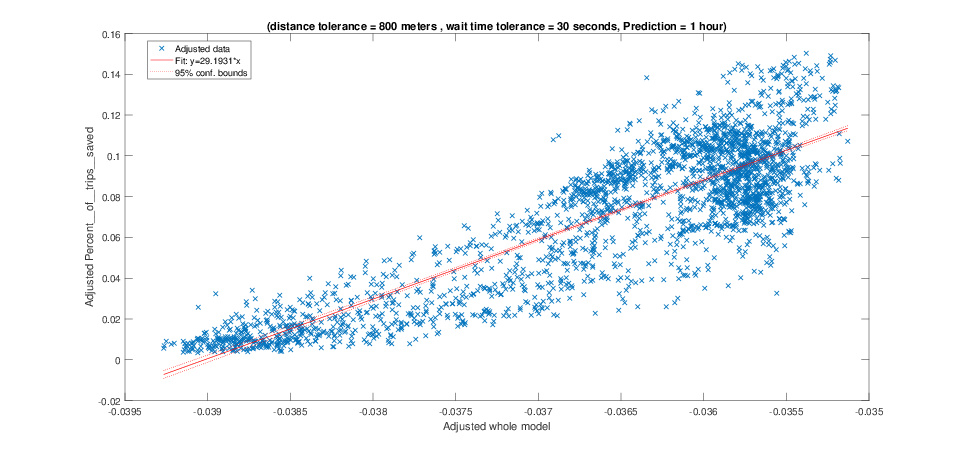}
\includegraphics[scale=0.4]{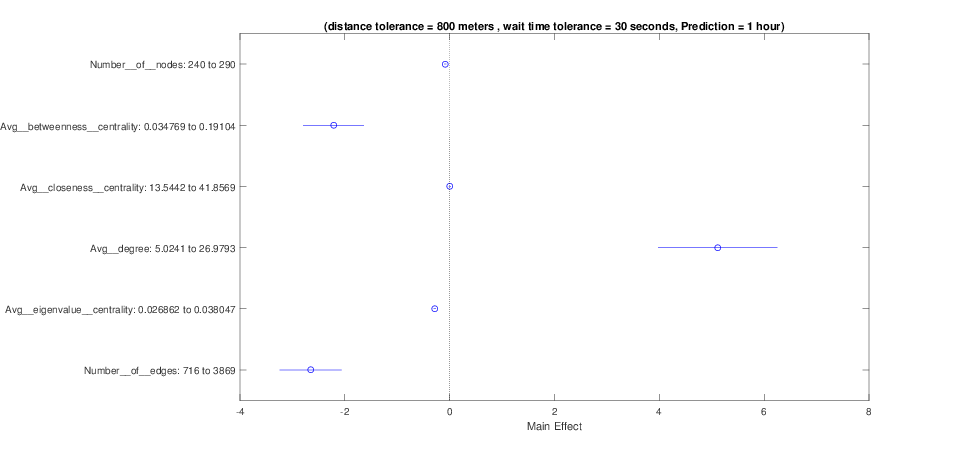}
\includegraphics[bb=0 0 850 700, clip=true, scale=0.35]{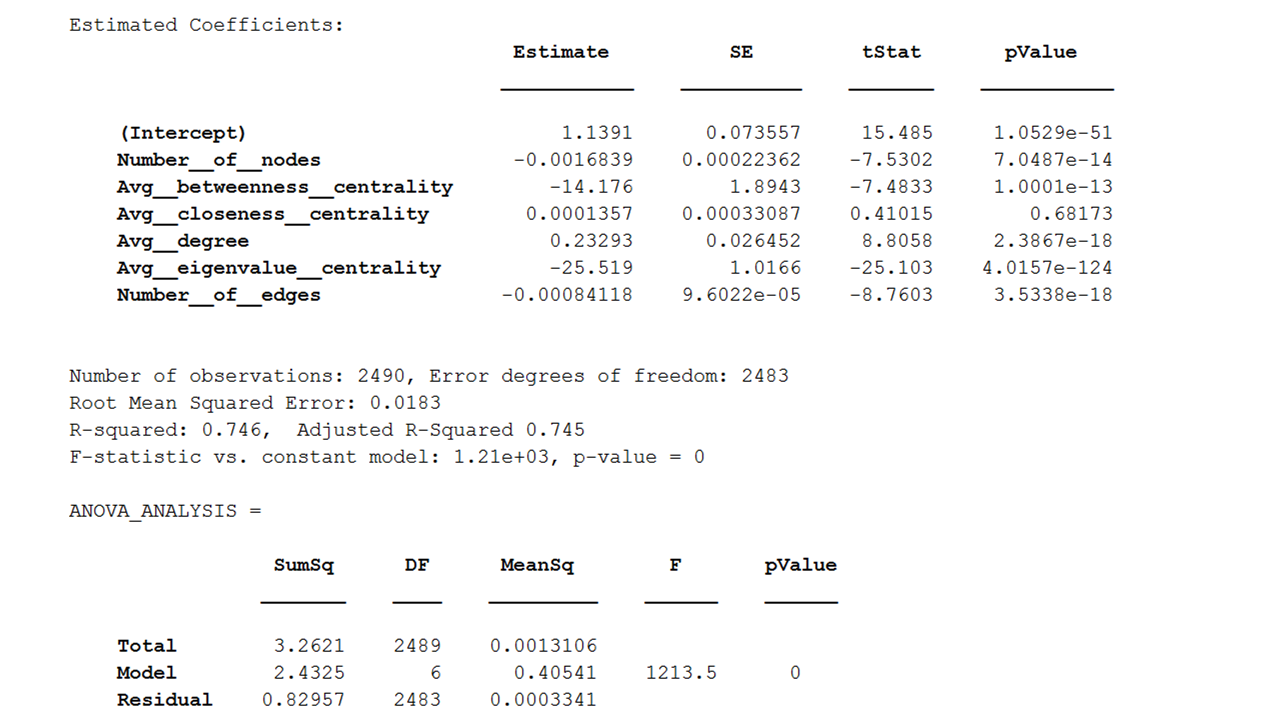}
\caption{Multilinear regression model, predicting the ride-sharing utilization using the dynamic network's properties 11.}
\label{fig:multilinear11}
\end{figure}

\begin{figure}[htbp]
\centering
\includegraphics[scale=0.4]{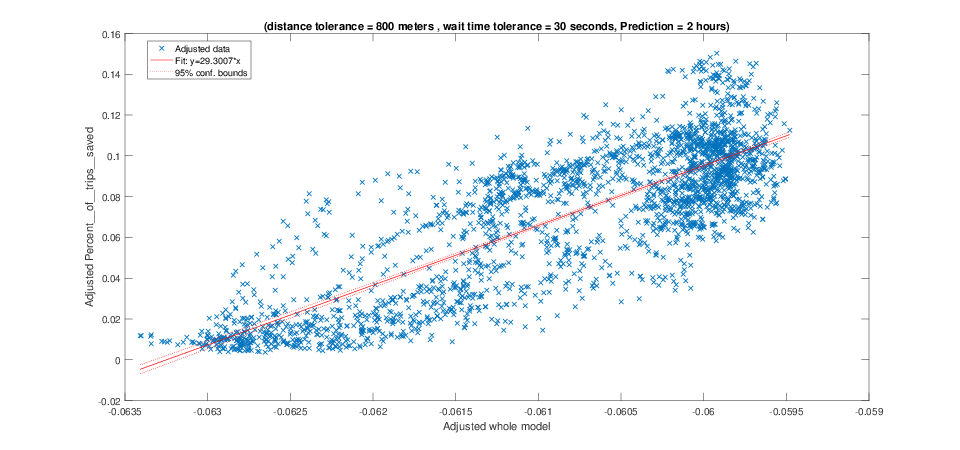}
\includegraphics[scale=0.4]{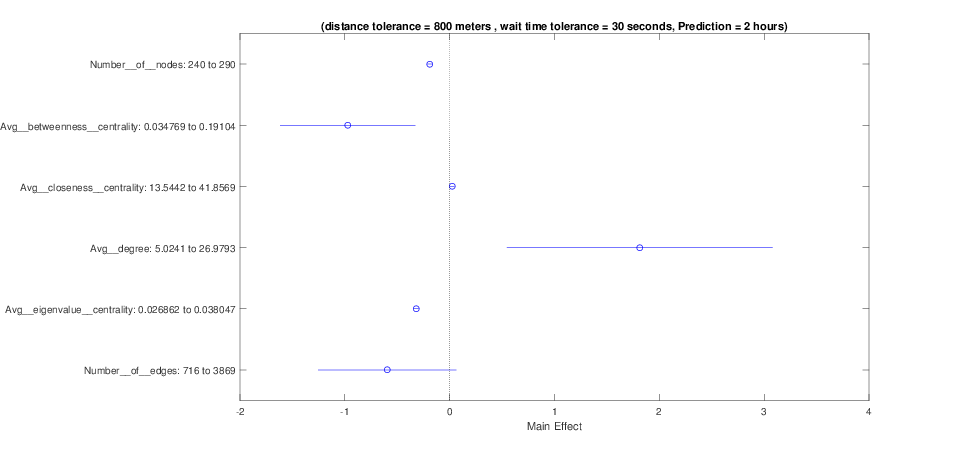}
\includegraphics[bb=0 0 850 700, clip=true, scale=0.35]{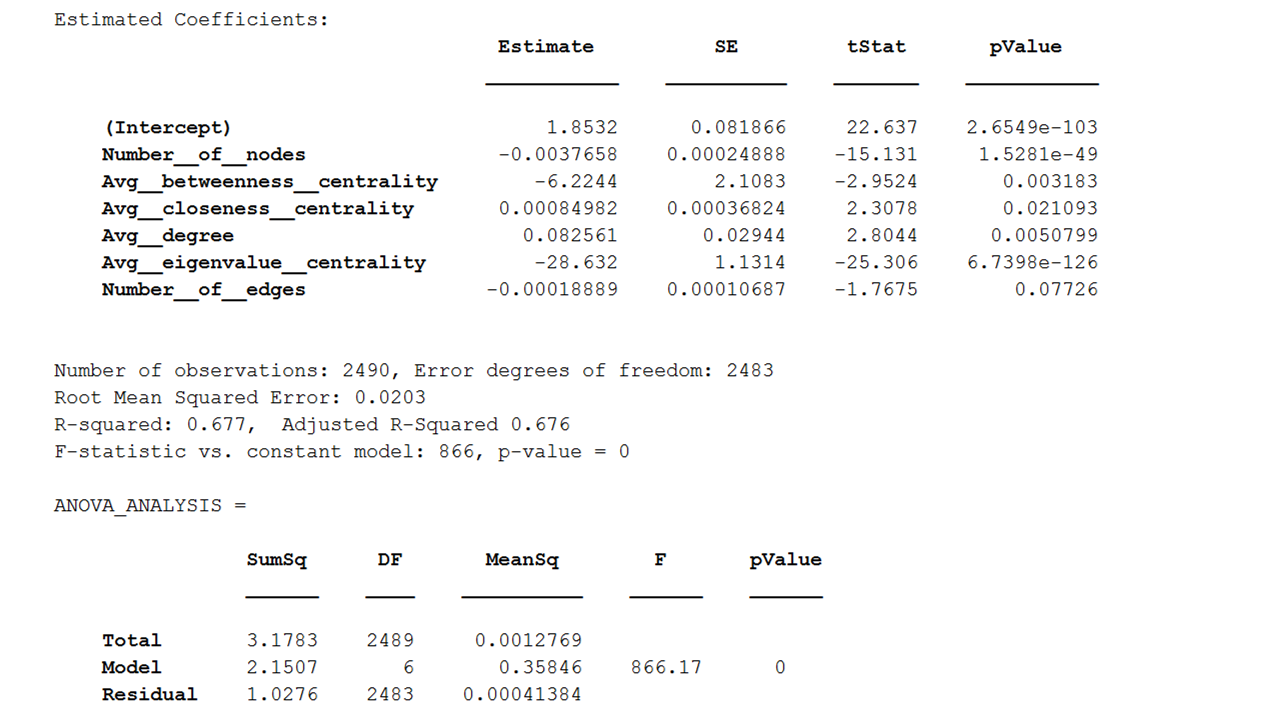}
\caption{Multilinear regression model, predicting the ride-sharing utilization using the dynamic network's properties 12.}
\label{fig:multilinear12}
\end{figure}

\begin{figure}[htbp]
\centering
\includegraphics[scale=0.4]{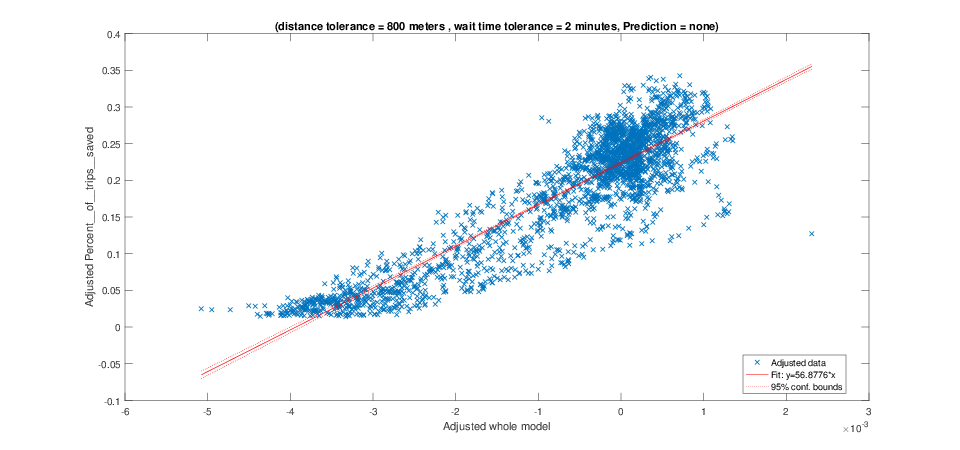}
\includegraphics[scale=0.4]{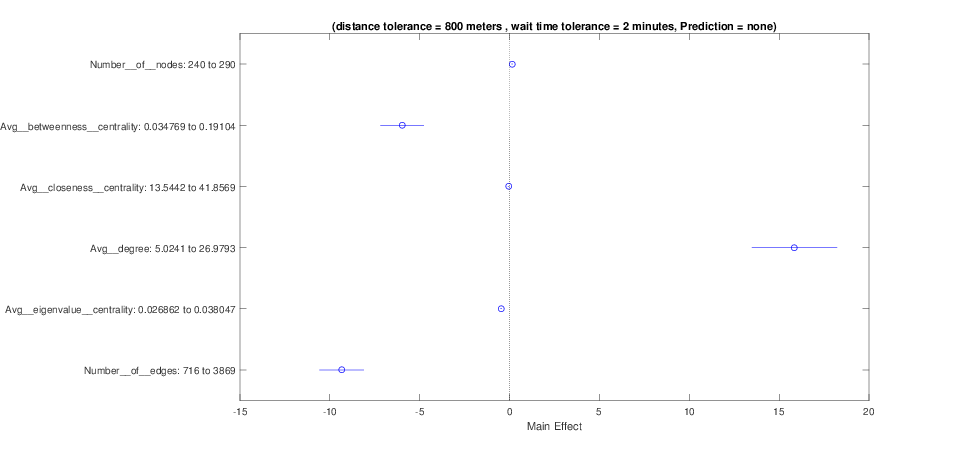}
\includegraphics[bb=0 0 850 700, clip=true, scale=0.35]{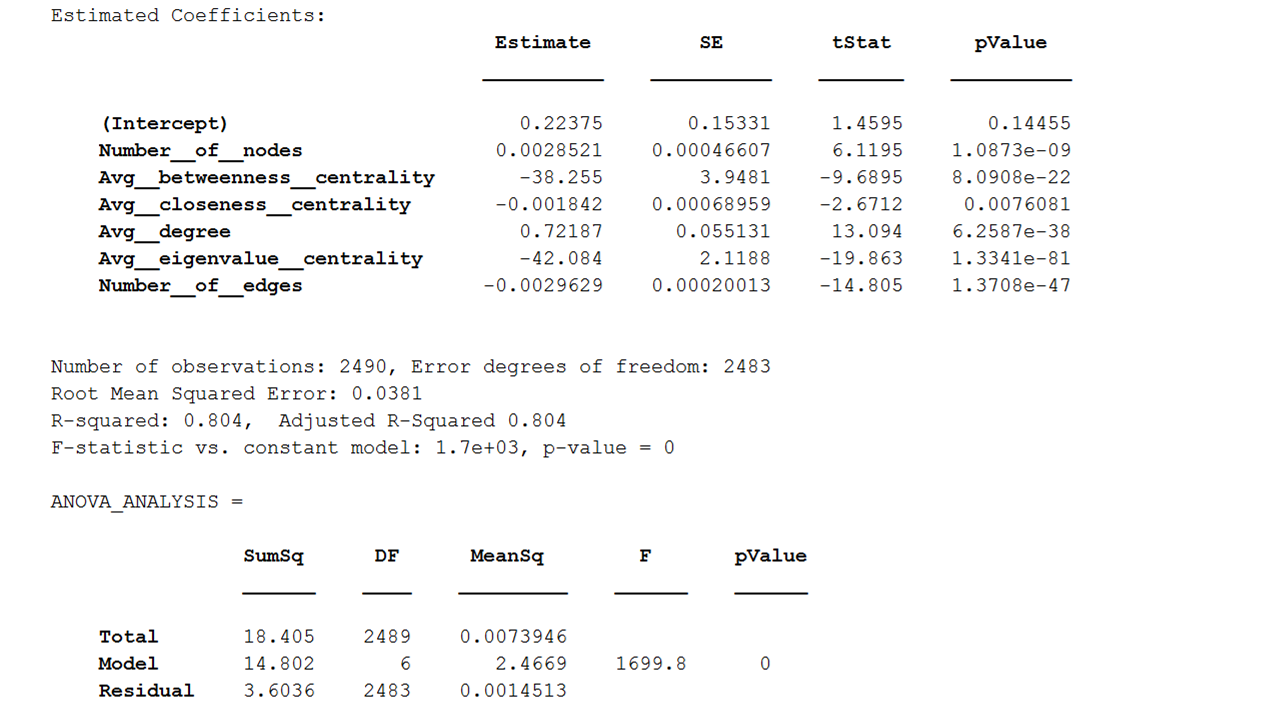}
\caption{Multilinear regression model, predicting the ride-sharing utilization using the dynamic network's properties 13.}
\label{fig:multilinear13}
\end{figure}

\begin{figure}[htbp]
\centering
\includegraphics[scale=0.4]{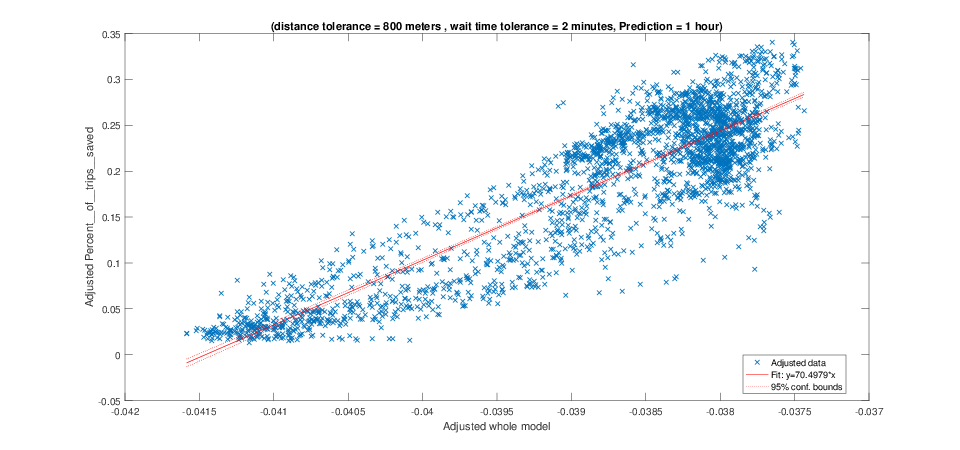}
\includegraphics[scale=0.4]{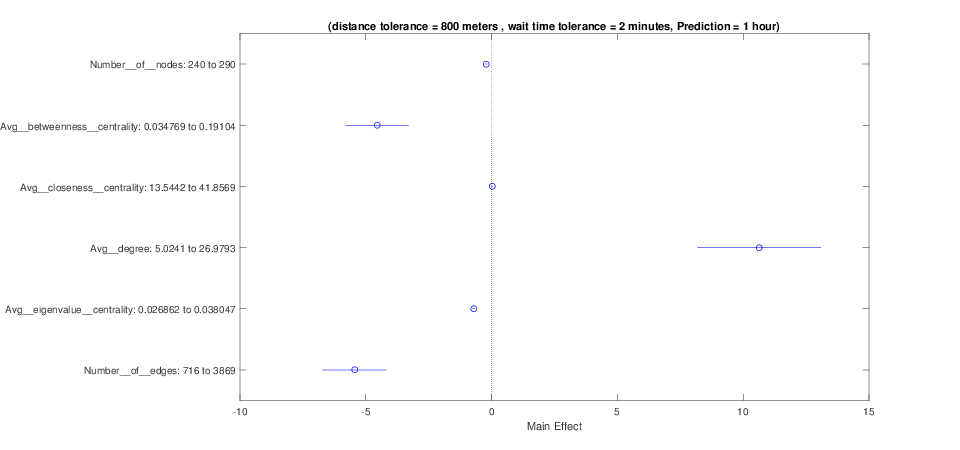}
\includegraphics[bb=0 0 850 700, clip=true, scale=0.35]{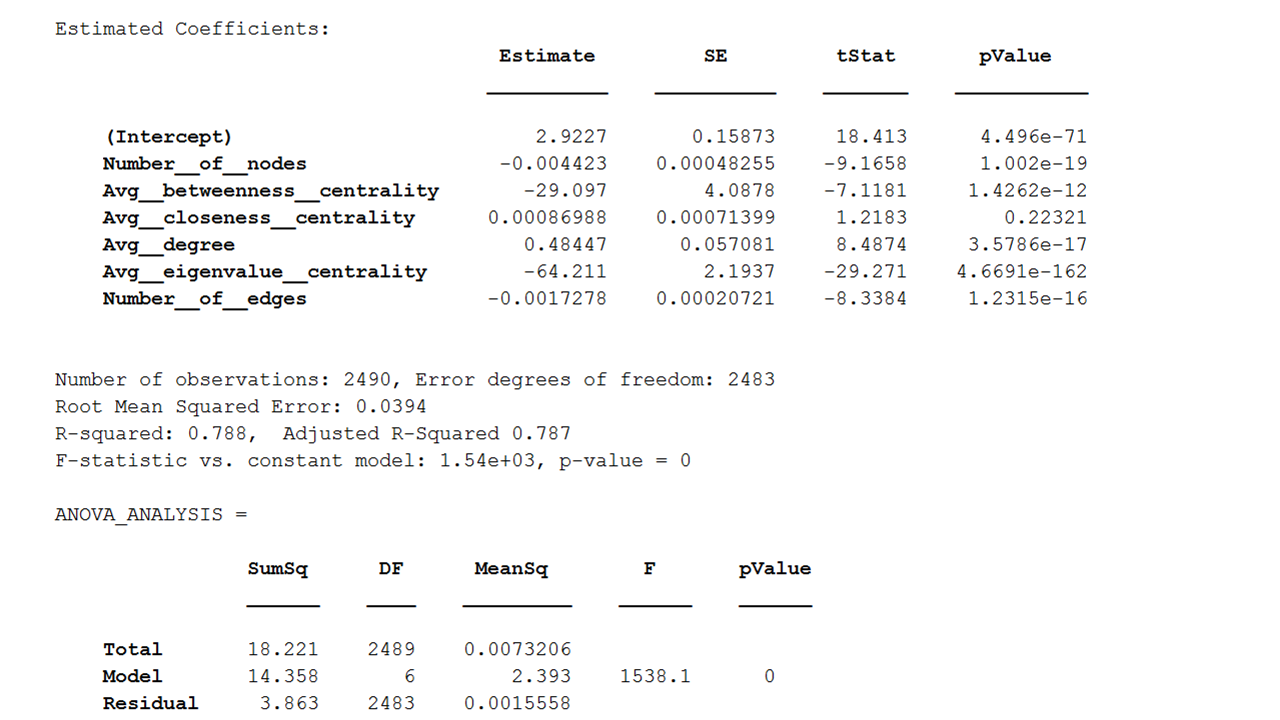}
\caption{Multilinear regression model, predicting the ride-sharing utilization using the dynamic network's properties 14.}
\label{fig:multilinear14}
\end{figure}

\begin{figure}[htbp]
\centering
\includegraphics[scale=0.4]{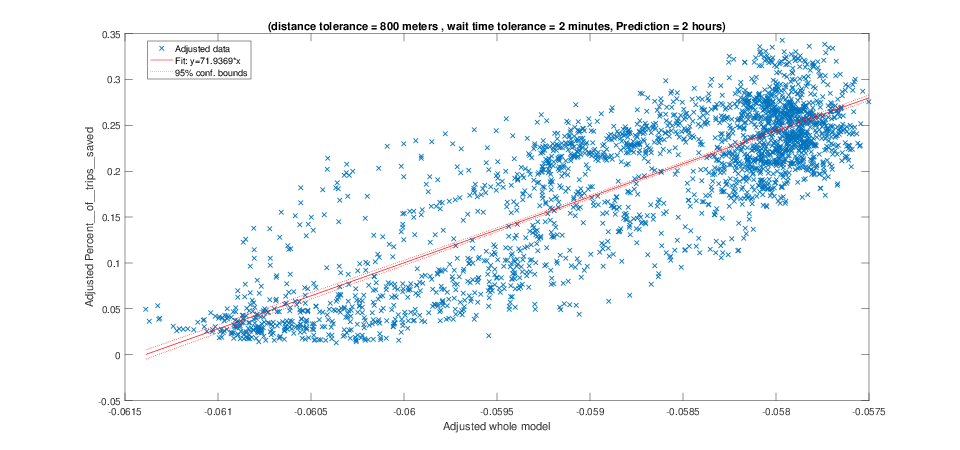}
\includegraphics[scale=0.4]{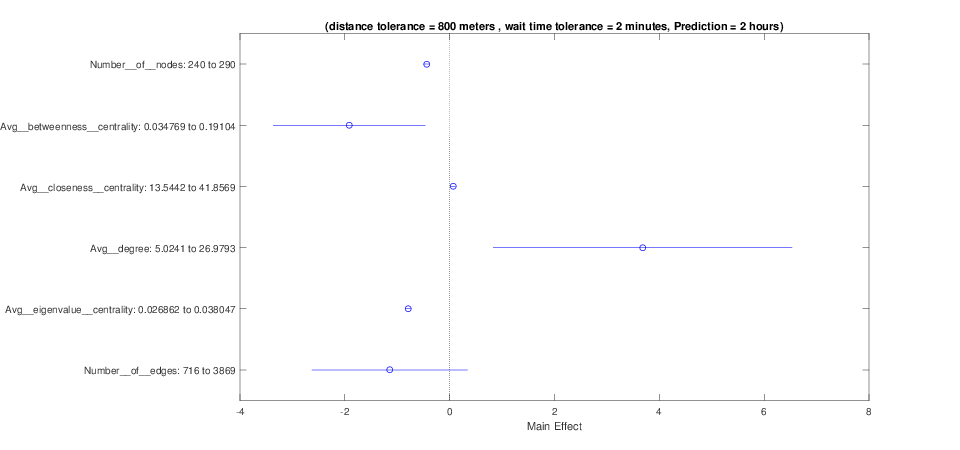}
\includegraphics[bb=0 0 850 700, clip=true, scale=0.35]{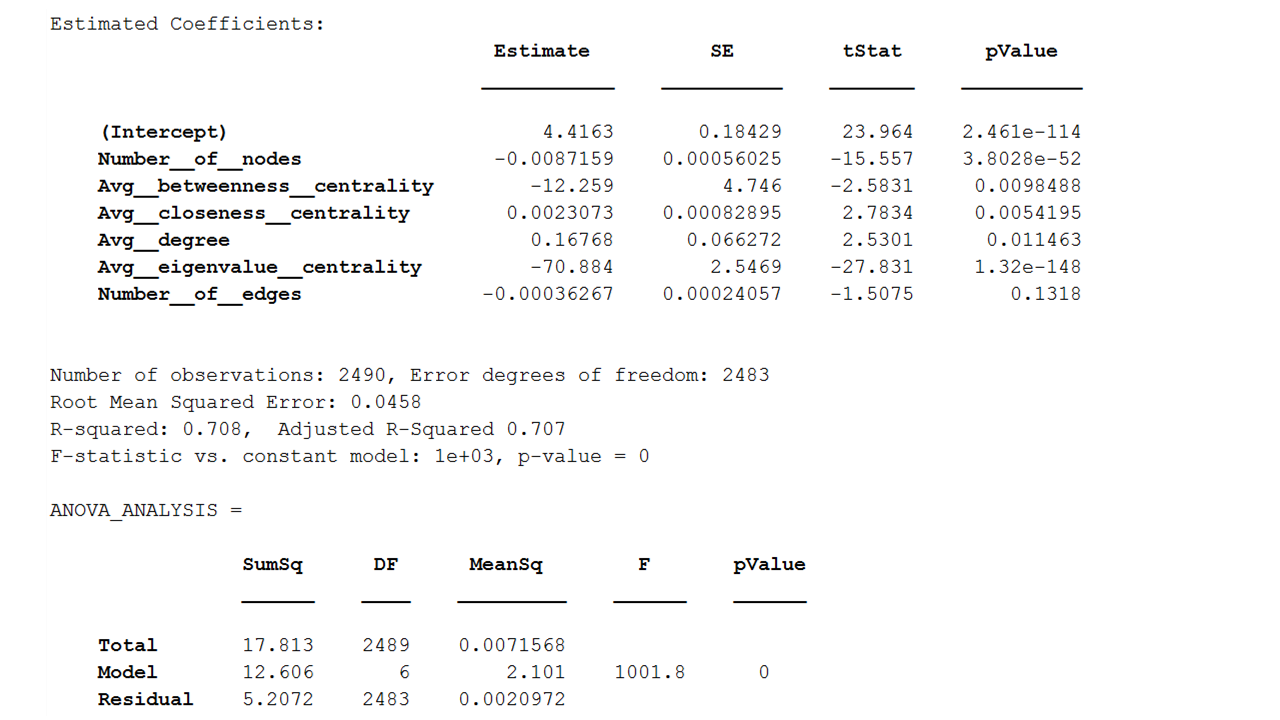}
\caption{Multilinear regression model, predicting the ride-sharing utilization using the dynamic network's properties 15.}
\label{fig:multilinear15}
\end{figure}

\begin{figure}[htbp]
\centering
\includegraphics[scale=0.4]{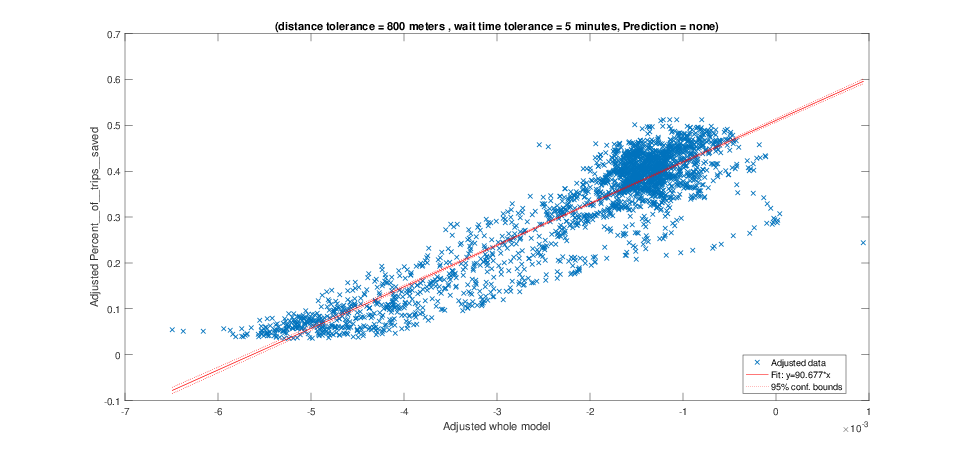}
\includegraphics[scale=0.4]{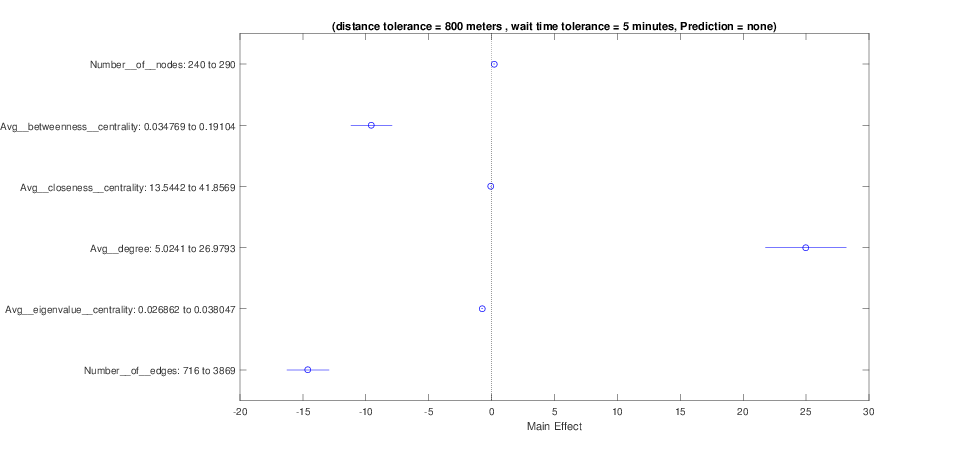}
\includegraphics[bb=0 0 850 700, clip=true, scale=0.35]{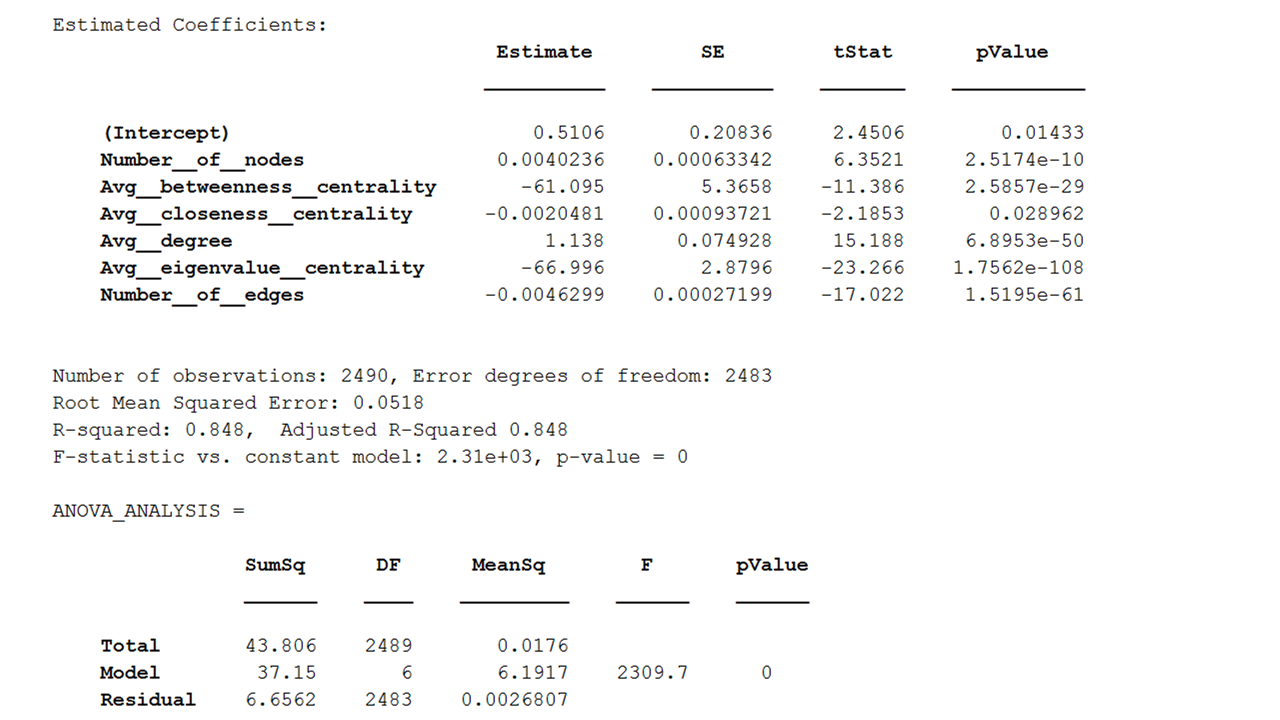}
\caption{Multilinear regression model, predicting the ride-sharing utilization using the dynamic network's properties 16.}
\label{fig:multilinear16}
\end{figure}

\begin{figure}[htbp]
\centering
\includegraphics[scale=0.4]{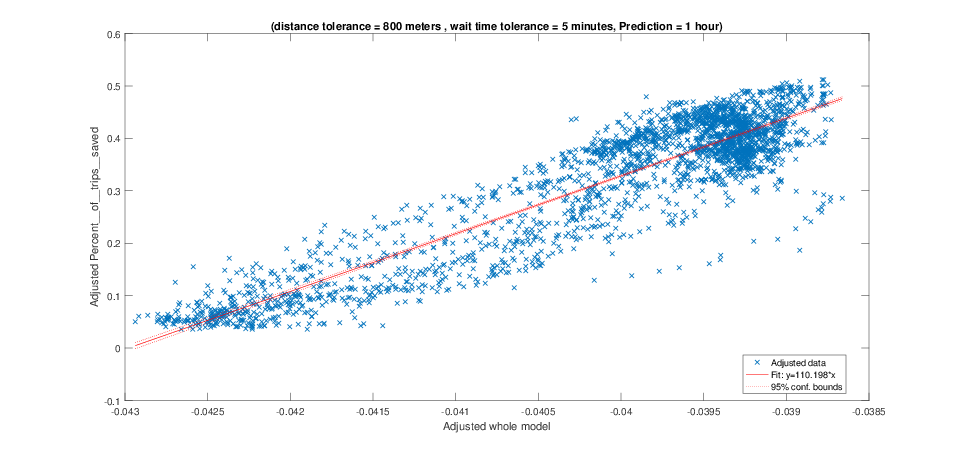}
\includegraphics[scale=0.4]{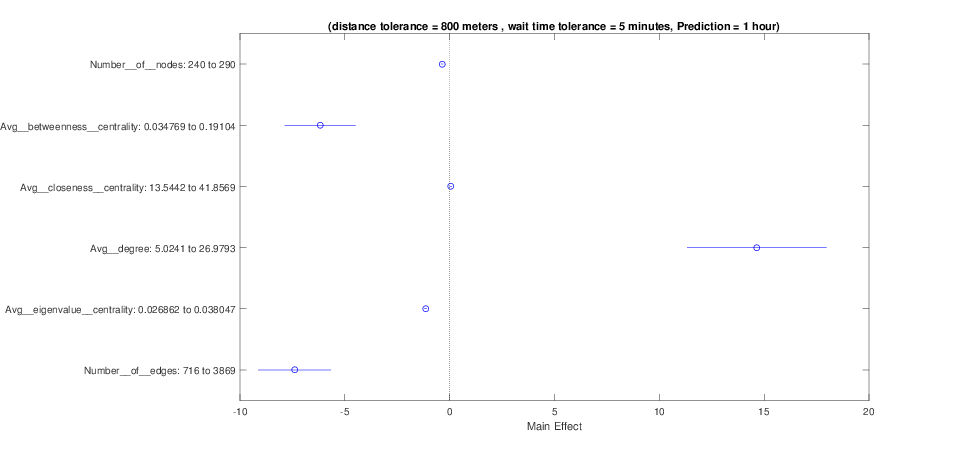}
\includegraphics[bb=0 0 850 700, clip=true, scale=0.35]{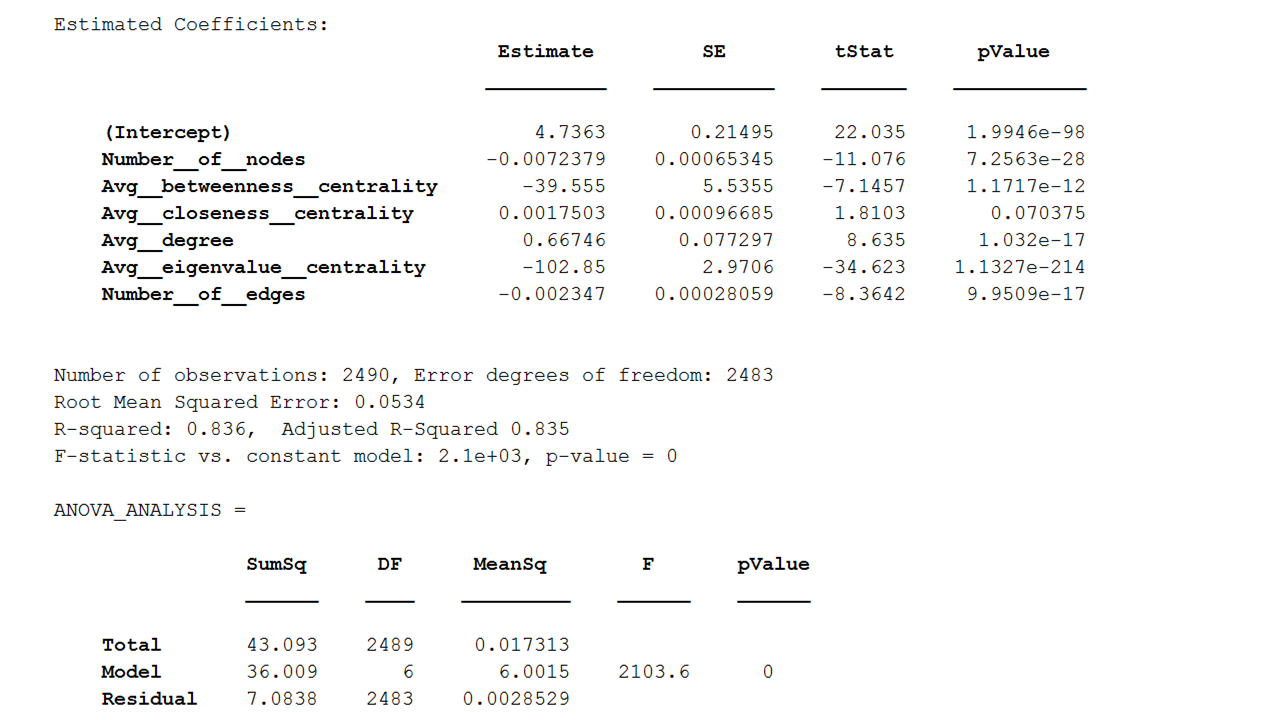}
\caption{Multilinear regression model, predicting the ride-sharing utilization using the dynamic network's properties 17.}
\label{fig:multilinear17}
\end{figure}

\begin{figure}[htbp]
\centering
\includegraphics[scale=0.4]{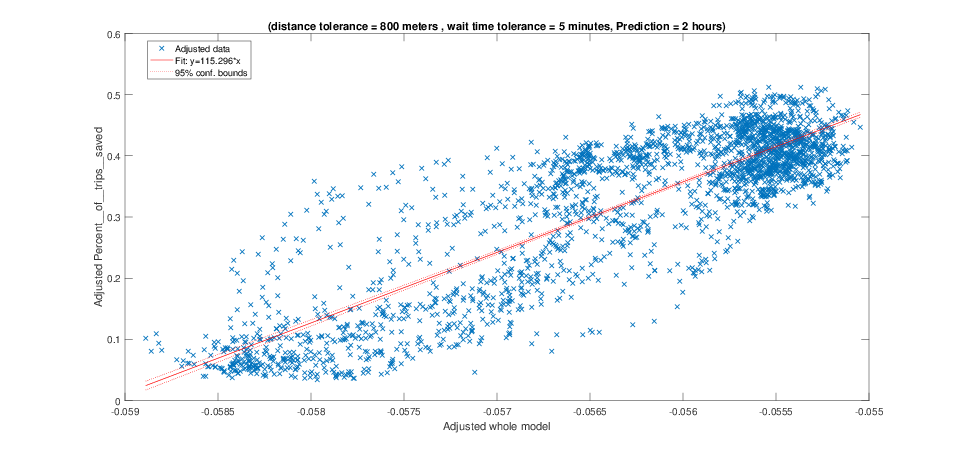}
\includegraphics[scale=0.4]{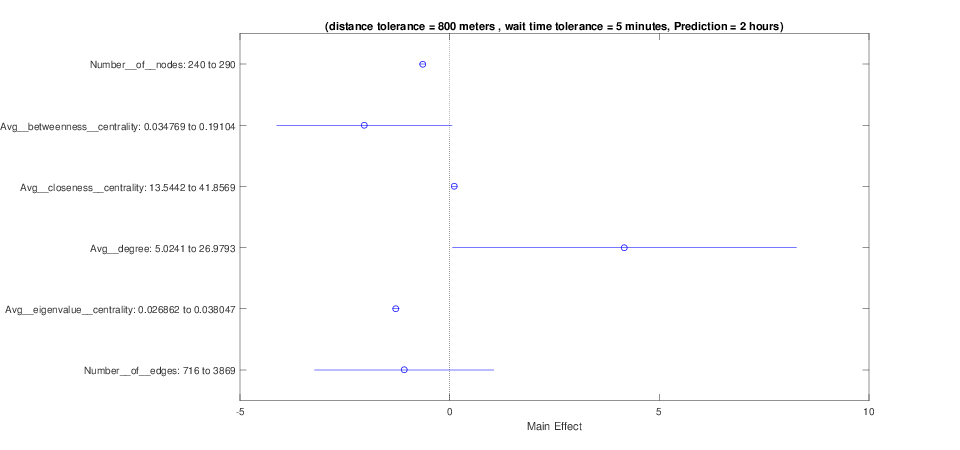}
\includegraphics[bb=0 0 850 700, clip=true, scale=0.35]{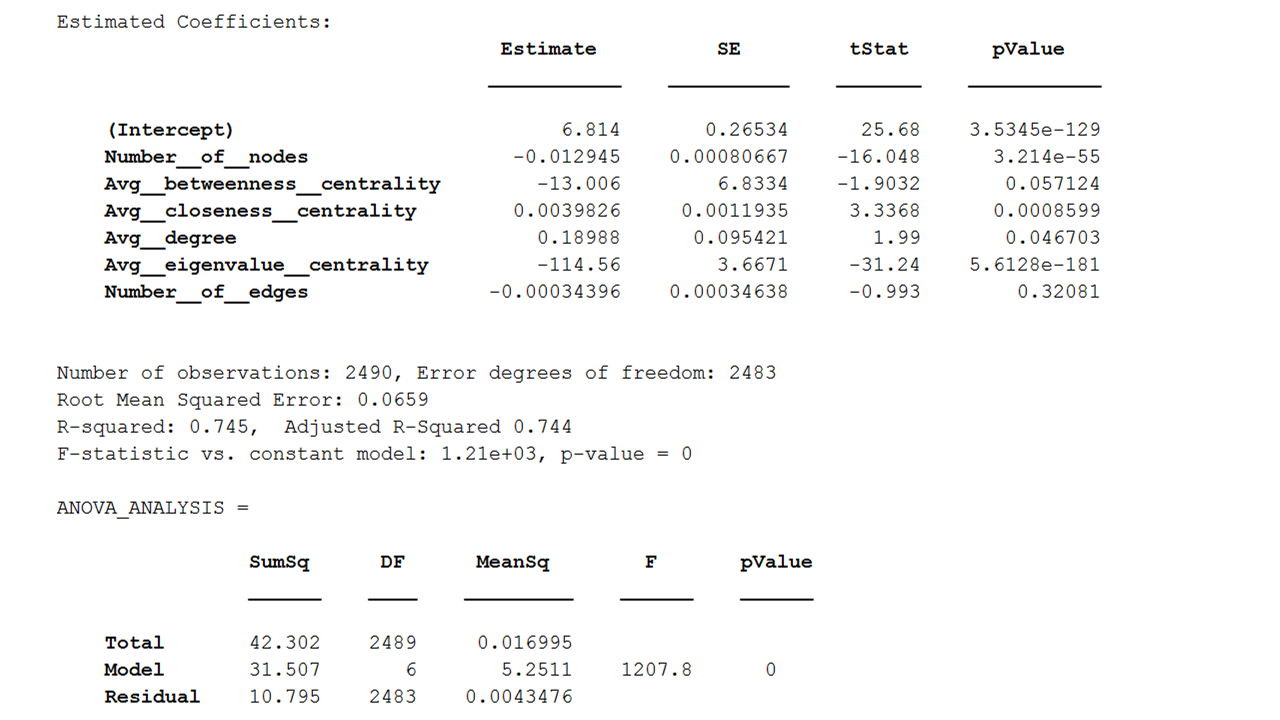}
\caption{Multilinear regression model, predicting the ride-sharing utilization using the dynamic network's properties 18.}
\label{fig:multilinear18}
\end{figure}

\clearpage


\begin{figure}[htbp]
\centering
\includegraphics[scale=0.5]{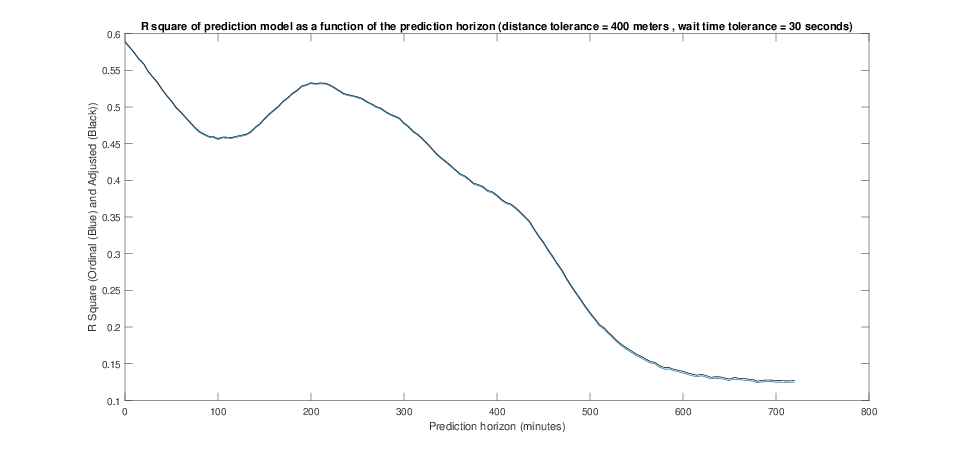}
\caption{The accuracy of the multilinear regression model, as measured by its $R^{2}$, as a function of the prediction horizon 1.}
\label{fig:multilinear_squarer1}
\end{figure}

\begin{figure}[htbp]
\centering
\includegraphics[scale=0.5]{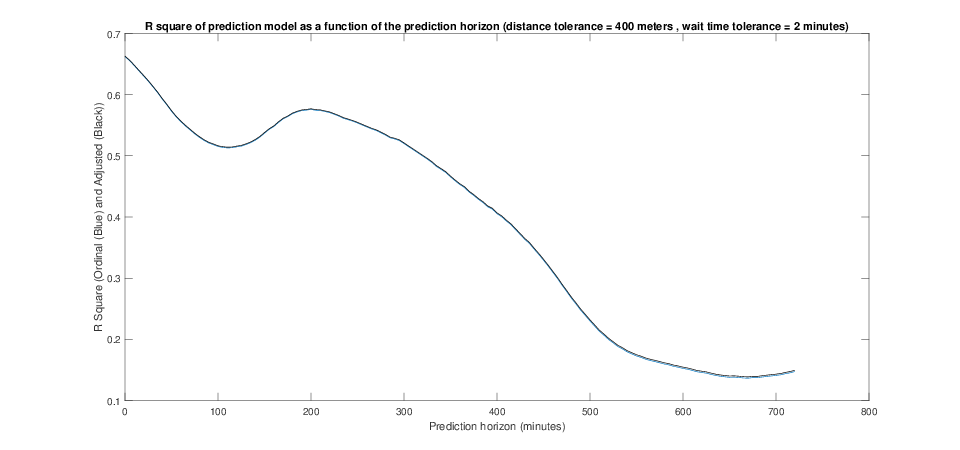}
\caption{The accuracy of the multilinear regression model, as measured by its $R^{2}$, as a function of the prediction horizon 2.}
\label{fig:multilinear_squarer2}
\end{figure}

\begin{figure}[htbp]
\centering
\includegraphics[scale=0.5]{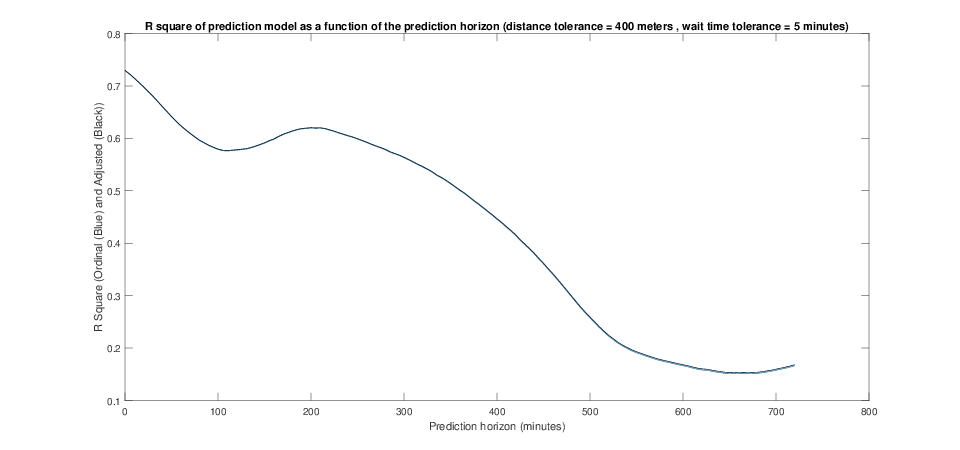}
\caption{The accuracy of the multilinear regression model, as measured by its $R^{2}$, as a function of the prediction horizon 3.}
\label{fig:multilinear_squarer3}
\end{figure}

\begin{figure}[htbp]
\centering
\includegraphics[scale=0.5]{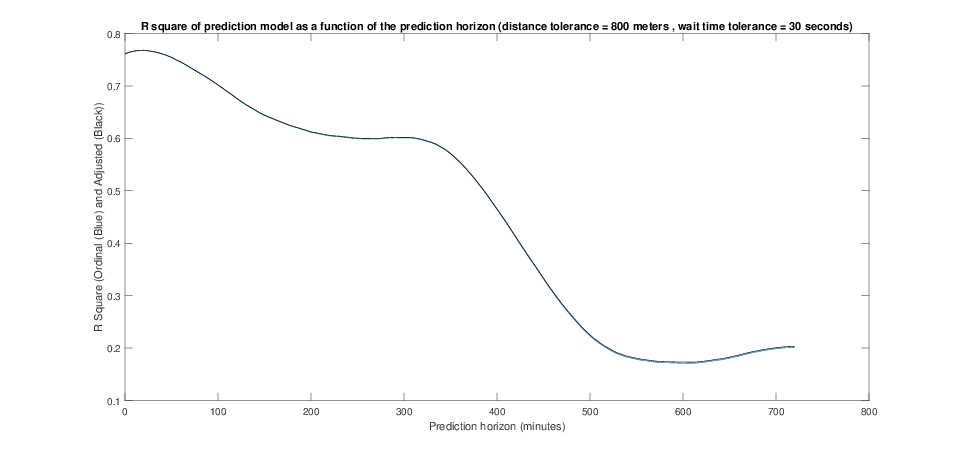}
\caption{The accuracy of the multilinear regression model, as measured by its $R^{2}$, as a function of the prediction horizon 4.}
\label{fig:multilinear_squarer4}
\end{figure}

\begin{figure}[htbp]
\centering
\includegraphics[scale=0.5]{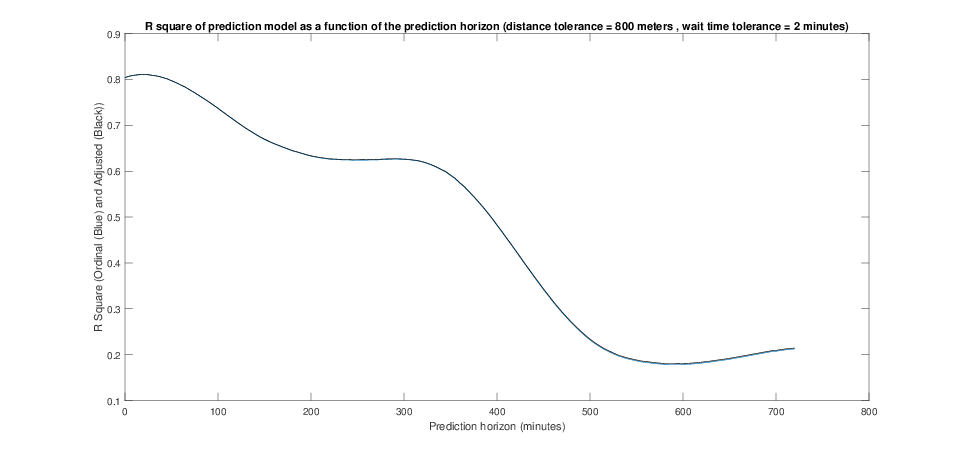}
\caption{The accuracy of the multilinear regression model, as measured by its $R^{2}$, as a function of the prediction horizon 5.}
\label{fig:multilinear_squarer5}
\end{figure}

\begin{figure}[htbp]
\centering
\includegraphics[scale=0.5]{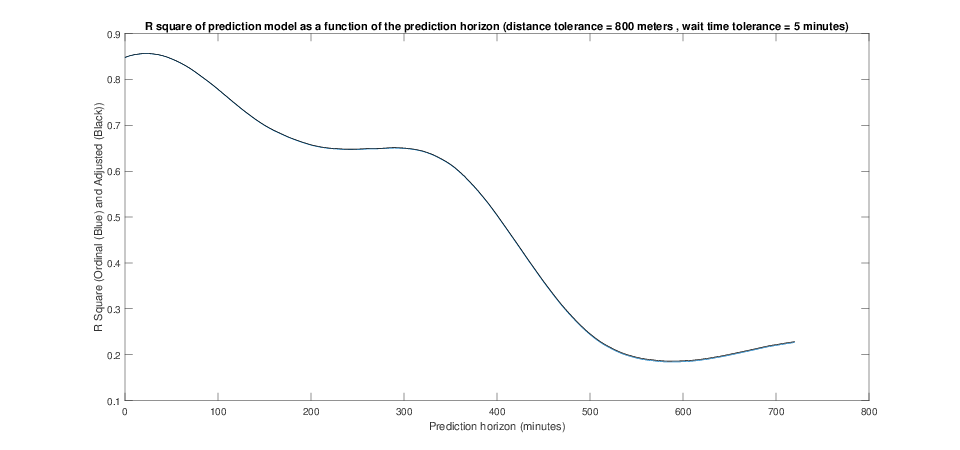}
\caption{The accuracy of the multilinear regression model, as measured by its $R^{2}$, as a function of the prediction horizon 6.}
\label{fig:multilinear_squarer6}
\end{figure}

\end{document}